\documentclass[journal]{IEEEtran}
\usepackage{multirow}
\usepackage{arydshln}
\usepackage{makecell}
\usepackage{threeparttable}

\usepackage{cite}

\ifCLASSINFOpdf
   \usepackage[pdftex]{graphicx}
   \graphicspath{{figures/}}
   \DeclareGraphicsExtensions{.pdf,.jpeg,.png}
\else
\fi
\usepackage{booktabs}
\usepackage{float}
\usepackage{amsmath}
\usepackage{amsthm}
\usepackage{amssymb}
\usepackage{algorithm}
\usepackage{algorithmic}
\usepackage{color}
\usepackage{verbatim}

\usepackage{pifont}

\DeclareMathOperator*{\argmax}{arg\,max}

\newcommand{\conti}{\textcolor{blue}{C}}
\newcommand{\discr}{\textcolor{magenta}{D}}

\newcommand{\cmark}{\textcolor{black}{\ding{51}}}
\newcommand{\xmark}{\textcolor{black}{\ding{55}}}

\begin{document}

%

\title{A Review of Bayesian Methods in Electronic Design Automation}
%
%
%

\author{Zhengqi Gao and Duane S. Boning,~\IEEEmembership{Fellow,~IEEE}
\thanks{Zhengqi Gao and Duane S. Boning are with the Department
of Electrical Engineering and Computer Science, Massachusetts Institute of Technology, Cambridge, MA 02139, USA (e-mail: \{zhengqi, boning\}@mit.edu). Zhengqi Gao is the corresponding author. We welcome comments and feedback on any aspect of our work, and please direct them to zhengqi@mit.edu.}
}

%
%

\markboth{Manuscript}%
{leave blank}
%



\maketitle

\begin{abstract}
The utilization of Bayesian methods has been widely acknowledged as a viable solution for tackling various challenges in electronic integrated circuit (IC) design under stochastic process variation, including circuit performance modeling, yield/failure rate estimation, and circuit optimization. As the post-Moore era brings about new technologies (such as silicon photonics and quantum circuits), many of the associated issues there are similar to those encountered in electronic IC design and can be addressed using Bayesian methods. Motivated by this observation, we present a comprehensive review of Bayesian methods in electronic design automation (EDA). By doing so, we hope to equip researchers and designers with the ability to apply Bayesian methods in solving stochastic problems in electronic circuits and beyond.
\end{abstract}


%
\IEEEpeerreviewmaketitle

\section{A Historical Introduction}

From an historical perspective, human efforts devoted to integrating electronic circuits onto a single chip date back to the early 20th century~\cite{bornIC_kilby1959miniaturized,bornIC_noyce1959semiconductor}. Accompanying and catalyzing the blossoming of electronic integrated circuits (ICs) is a methodology termed electronic design automation (EDA)~\cite{revieweda_macmillen2000industrial,revieweda_wang2009electronic,revieweda_lavagno2016electronic,revieweda_breuer1981survey,revieweda_darringer2000eda}. In layman's terms, EDA attempts to use algorithms to enable or speed up the design of an electronic chip while reducing human workloads at the same time. EDA is present at each abstraction level in the modern electronic IC design flow~\cite{layout_baker2019cmos}, such as efficient circuit simulation~\cite{simulation_pederson1984historical,simulation_najm2010circuit,simulation_ogrodzki2018circuit,simulation_rewienski2003trajectory,simulation_tan2007advanced} at the front-end, and automatic circuit layout~\cite{layout_barahona1988application,layout_lengauer2012combinatorial,layout_alam2002comprehensive} at the back-end.

The research focus of the EDA community has changed alongside, as well as reflected, the development of electronic ICs. In the early stages from around 1900 to 1950 when the electronic IC industry was at its infancy, most EDA research efforts (in a broad sense) were oriented towards understanding semiconductor physics~\cite{semphysics_dekker1957solid} and simulating an electronic circuit~\cite{simulation_nagel1996life}. A few prominent and influential theories were proposed, such as modified nodal analysis~\cite{mna_ho1975modified}, Tellegen's theorem~\cite{tellegen_penfield1970generalized} and adjoint methods~\cite{adjoint_director1969generalized,adjoint_director1969automated}, being accompanied by a few successful demonstrations of bipolar~\cite{bipolar_kroemer1982heterostructure} and field effect transistors~\cite{fft_shockley1952unipolar,fft_dacey1955field}.

Built upon this solid foundation, the period from 1950 to 2000 marked the prime time for both electronic ICs and EDA. In this period, electronic ICs experienced tremendous development, and many products (e.g., digital television, handheld mobile phones) marched into our daily life. Such exceptional accomplishments were not possible without the assistance of EDA, during which many EDA companies were founded (e.g., Cadence, Synopsys, and many others), multiple research subareas were formulated (e.g., logic synthesis~\cite{logicsynthesis_devadas1994logic,logicsynthesis_rudell1989logic,logicsynthesis_brayton1990multilevel,logicsynthesis_sasao1993logic}, timing analysis~\cite{timing_agarwal2005circuit,timing_pillage1990asymptotic}, fault diagnosis~\cite{fault_slamani1992analog,fault_spina1997linear}, and functional/formal verification~\cite{formal_coudert2003unified,formal_drechsler2004advanced}), and many tools were developed and standardized (e.g., Verilog~\cite{verilog_palnitkar2003verilog,verilog_thomas2008verilog} and VHDL~\cite{vhdl_navabi1997vhdl,vhdl_ashenden2010designer}).

Later, the arrival of the new millennium brought fresh problems to the electronic IC industry. MOSFET technology, for the first time, scaled down to the deep submicron regime (i.e., feature size less than 100~nm). Denser integration could be achieved on a single chip~\cite{vlsi_sherwani2012algorithms}, while at the same time, the impact of manufacturing process variation became significant as the MOSFET technology node continued to shrink~\cite{dfm_boning1999models,dfm_borkar2005designing,dfm_sarangi2008varius,dfm_Michael2008design,dfm_kuhn2011process}. These stochastic process variations (e.g., variation of transistor length or width) could induce severe circuit performance degradation and reduce the product yield. To this end, many research efforts have been devoted to effectively modeling the impact of process variation~\cite{dfm_agarwal2007characterizing,dfm_mutlu2005statistical}, hence further improving the process itself.

For this goal, Bayesian methods were introduced to the EDA community around the 2000s~\cite{fault_liu2006parametric,perfmodeling_bhanja2003switching,perfmodel_firstbmf_li2013bayesian,perfmodel_nottypical_hsu2007forecasting,fault_mittelstadt1995application,fault_brandt1997circuit,processcontrol_rao1996monitoring,processcontrol_yousry1991process}, due to their capabilities to handle uncertainty. The core of Bayesian approaches relies on Bayes' theorem, where people encode their knowledge about the problem in a \emph{prior distribution}, and then a \emph{posterior distribution} is derived from it after observing some data~\cite{bayestheory_bernardo2009bayesian,bayestheory_bishop2006pattern}. Since its introduction, we have witnessed a large number of works successfully applying Bayesian methods to address stochastic problems, initially in circuit modeling and analysis under process variation~\cite{perfmodeling_bhanja2003switching,perfmodel_nottypical_hsu2007forecasting,perfmodel_zhang2010toward,perfmodel_bmf_li2012efficient,perfmodel_bmf_wang2013bayesian,perfmodel_firstbmf_li2013bayesian,perfmodel_selfhealing_sun2013indirect,perfmodel_moment_gu2013efficient,perfmodel_mp_li2014efficient,perfmodel_clbmf_wang2015co,perfmodel_huang2015efficient,perfmodel_dualpriorbmf_huang2016efficient,perfmodel_cbmf_wang2016correlated,perfmodel_tao2019large,perfmodel_liu2020transfer,perfmodel_cbmf_gao2022correlated}, and later in circuit optimization~\cite{bocircuitopt_liu2014gaussian,bocircuitopt_lyu2017efficient,bocircuitopt_lyu2018multi,bocircuitopt_lyu2018batch,bocircuitopt_zhang2019bayesian,bocircuitopt_zhang2019efficient,bocircuitopt_gao2019efficient,bocircuitopt_pan2019analog,bocircuitopt_zhang2020efficient,bocircuitopt_he2020efficient,bocircuitopt_abdelaal2020bayesian,bocircuitopt_lu2020mixed,bocircuitopt_touloupas2021locomobo,bocircuitopt_touloupas2021local,bocircuitopt_liu2021parasitic,bocircuitopt_zhang2021efficient,bocircuitopt_liao2021high,bocircuitopt_huang2021bayesian,bocircuitopt_lu2021automated,bocircuitopt_huang2021robust,bocircuitopt_wang2021high,bocircuitopt_touloupas2021optimization,bocircuitopt_touloupas2022mixed,bocircuitopt_chen2022high,bocircuitopt_he2022batched,bocircuitopt_vicsan2022automated,bocircuitopt_touloupas2022mixed2,bocircuitopt_fu2022batch,bocircuitopt_wang2022analog,bocircuitopt_zhang2022lineasybo,bocircuitopt_huang2022bayesian,bocircuitopt_zhao2022novel,bocircuitopt_yin2022asynchronous,bocircuitopt_yin2022fast,bocircuitopt_yin2022efficient}.

At this time, we are entering a post-Moore era, and many new technologies are being investigated~\cite{quantumn_gu2017microwave,sp_wim2018silicon,mtj_goto2021high,cnt_hills2019modern}. Many of these, like electronic ICs twenty years ago, are facing problems due to process variation. For instance, there is a great deal of interest in building accurate stochastic compact device models and analyzing yield for integrated silicon photonics\cite{sp_wim2018silicon}, for which many of the Bayesian methods developed for electronic ICs can be adapted and applied. Inspired by this observation, we sense the urgent need to review what has been achieved using Bayesian methods in EDA, so that we can prepare ourselves when being challenged by similar questions in these emerging domains. We hope that with this review paper, researchers and designers without prior experience with Bayesian methods are able to apply them when facing stochastic problems in electronic circuits and other novel areas.

Our paper is organized as follows. In Section~\ref{sec:recap_bayes}, we present the basics of Bayesian methods, with an emphasis on those used frequently in EDA. In Section~\ref{sec:preliminary}, we introduce a framework for ease of later review and comment on Bayesian EDA. In Section~\ref{sec:modeling}, we describe how to apply Bayesian inference to solve modeling, analysis, and diagnosis tasks. Next, in Section~\ref{sec:optimization}, we highlight the recent trend of using Bayesian optimization to automate designs. Finally, we describe future possibilities in Section~\ref{sec:future} and conclude in Section~\ref{sec:conclusion}.

\section{Recap of Bayesian Methods}\label{sec:recap_bayes}
We explain Bayesian framework in the first subsection, followed by a description of Gaussian process regression and Bayesian optimization in the second subsection. These two subsections respectively provide the background needed for Section~\ref{sec:modeling} and~\ref{sec:optimization}.

\subsection{Bayes' Theorem and Bayesian Inference}

\begin{figure}[!htb]
    \centering
    \includegraphics[width=0.9\linewidth]{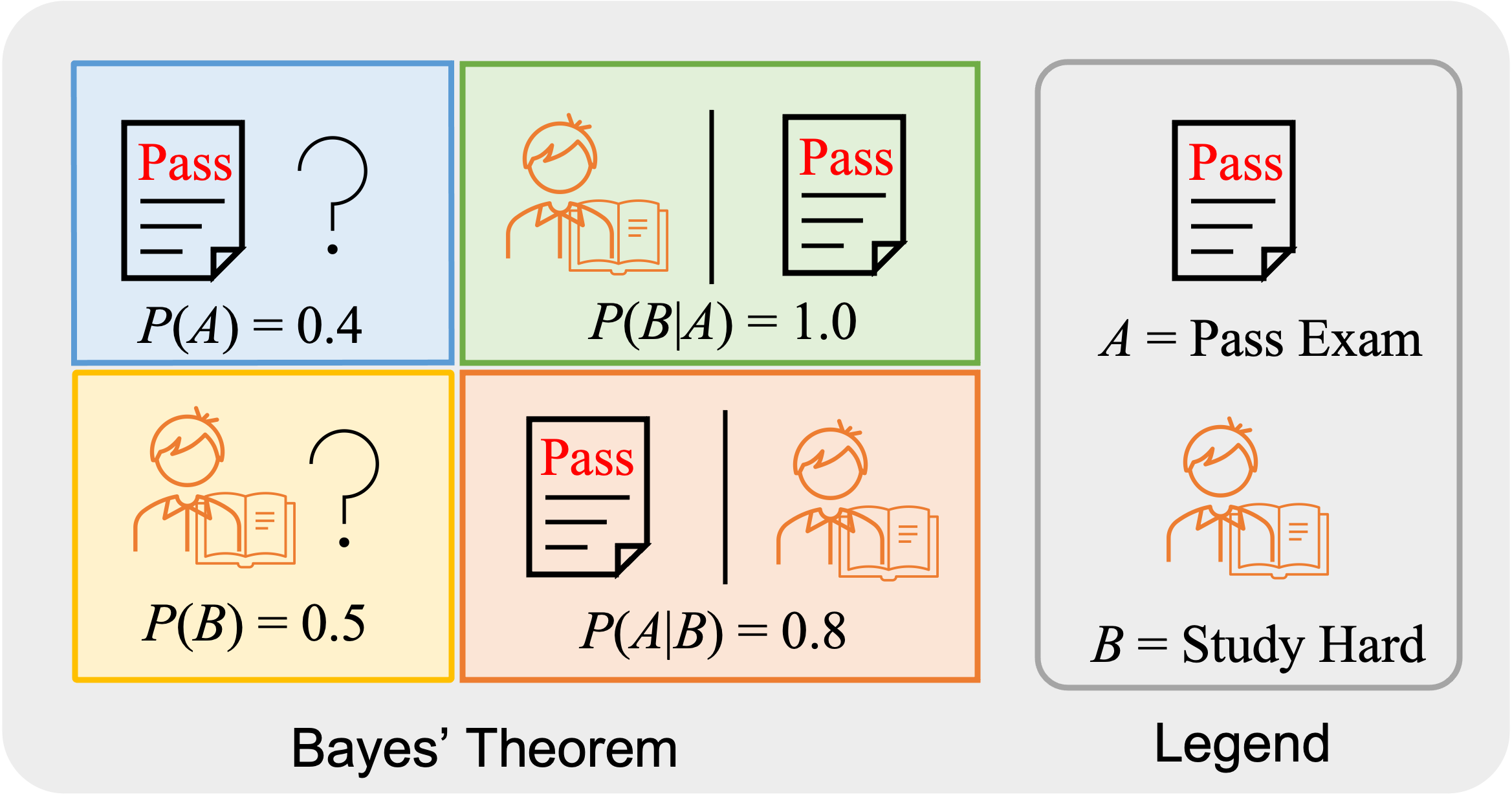}
    \caption{An example of Bayes' theorem. Observing event $B$ can change the probability of event $A$ happening from $P(A)=0.4$ to $P(A|B)=0.8$. Bayes' theorem states that knowing any three of the four probabilities determines the remaining one, but usually the conditional probability $P(A|B)$ or $P(B|A)$ is of interest.}
    \label{fig:bayes_theorem}
\end{figure}

Bayes' theorem states that for two events $A$ and $B$:

\begin{equation}\label{eq:original_bayes_theorem}
    P(A|B)=\frac{P(B|A)\,P(A)}{P(B)}
\end{equation}
where $P(A|B)$ denotes the probability of event $A$ occurring given that $B$ happens, and $P(A)$ means the probability of observing $A$ without any given conditions. When moving $P(B)$ from the denominator to the left hand side, we obtain the product rule of probability: $P(A|B)P(B)=P(B|A)P(A)$, and this product represents $P(A,B)$, i.e., the probability of event $A$ and $B$ happening simultaneously. Fig.~\ref{fig:bayes_theorem} shows an example of Bayes' theorem.



\begin{figure}[!htb]
    \centering
    \includegraphics[width=0.75\linewidth]{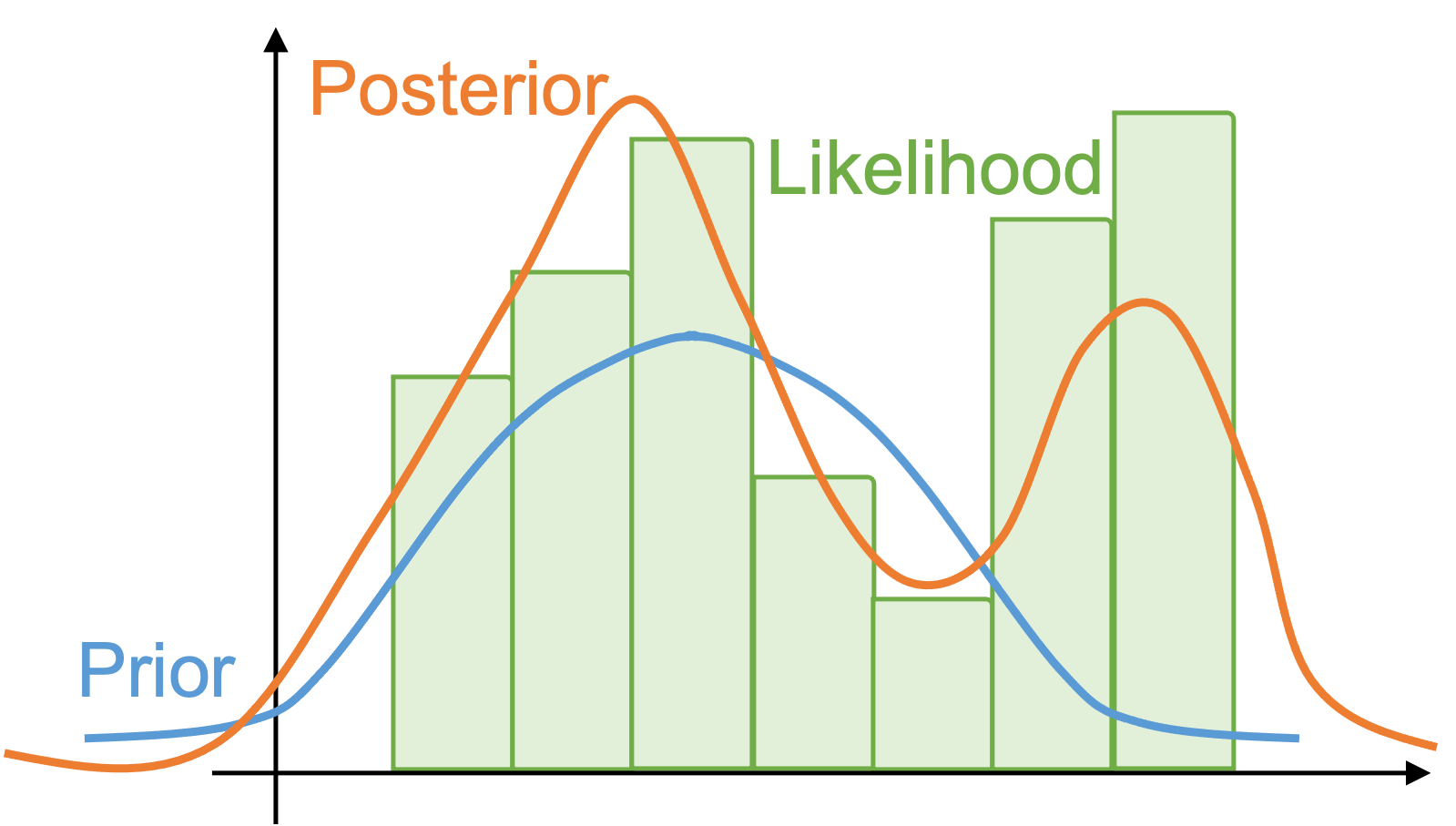}
    \caption{An illustration of Bayesian inference. Posterior is calibrated from the prior by using the likelihood. Colors are coded consistently as in Fig.~\ref{fig:bayes_theorem}.}
    \label{fig:bayesinference}
\end{figure}

Built upon Bayes' theorem, the framework of Bayesian inference has been proposed and widely used~\cite{bayestheory_bernardo2009bayesian,bayestheory_bishop2006pattern}. In many problems as demonstrated later, we are concerned with the distribution or the value (i.e., point estimate) of a model parameter $\boldsymbol{\theta}$ given a set of observed data $\mathcal{D}$. Such a distribution is denoted by $p(\boldsymbol{\theta}|\mathcal{D})$ and termed the \emph{posterior} distribution in Bayesian statistics. By analogy to Eq.~(\ref{eq:original_bayes_theorem}), we can calculate the posterior using:
\begin{equation}\label{eq:bayes_theorem1}
p(\boldsymbol{\theta}|\mathcal{D})=\frac{p(\mathcal{D}|\boldsymbol{\theta})\,p(\boldsymbol{\theta})}{p(\mathcal{D})}
\end{equation}
where $p(\boldsymbol{\theta})$ represents the distribution of $\boldsymbol{\theta}$ without any conditioning, and is called \emph{prior} distribution.  $p(\mathcal{D}|\boldsymbol{\theta})$ represents how probable the observed dataset $\mathcal{D}$ is given a specific $\boldsymbol{\theta}$, and is termed the \emph{likelihood} function~\cite{bayestheory_bishop2006pattern}. \footnote{We emphasize that $p(\mathcal{D}|\boldsymbol{\theta})$ is not a probability density function of $\boldsymbol{\theta}$. Namely, the integral of $p(\mathcal{D}|\boldsymbol{\theta})$ with respect to $\boldsymbol{\theta}$ is not necessarily~$1.$} Note that the denominator in Eq.~(\ref{eq:bayes_theorem1}) is not a function of $\boldsymbol{\theta}$, and thus, can be comprehended as a normalization constant ensuring that $p(\boldsymbol{\theta}|\mathcal{D})$ is a valid probability density function (i.e., the integral of $p(\boldsymbol{\theta}|\mathcal{D})$ in the domain of $\boldsymbol{\theta}$ equals $1$). To this end, Eq.~(\ref{eq:bayes_theorem1}) is usually written as:
\begin{equation}\label{eq:bayes_theorem2}
p(\boldsymbol{\theta}|\mathcal{D})\propto p(\mathcal{D}|\boldsymbol{\theta})\,p(\boldsymbol{\theta}) 
\end{equation}
in the Bayesian literature~\cite{bayestheory_bernardo2009bayesian,bayestheory_bishop2006pattern}. We emphasize that in Eq.~(\ref{eq:original_bayes_theorem}), the capitalized $P(\cdot)$ represents the probability of an event, while the uncapitalized $p(\cdot)$ in Eq.~(\ref{eq:bayes_theorem2}) represents a probability density function. Fig.~\ref{fig:bayesinference} demonstrates the concept of Bayesian inference.

We use Bayesian linear regression~\cite{bayestheory_bishop2006pattern} as a toy example to demonstrate how to utilize Eq.~(\ref{eq:bayes_theorem2}). Suppose a dataset $\mathcal{D}=\{(\mathbf{x}_i, y_i)|i=1,2,\cdots, N\}$ is given, where $\mathbf{x}\in\mathbb{R}^H$ and $y\in\mathbb{R}$, and we are asked to obtain a linear model parameterized by $\boldsymbol{\theta}$: $f_{\boldsymbol\theta}(\mathbf{x})=\boldsymbol{\theta}^T\mathbf{x}$, to fit the mapping from $\mathbf{x}$ to $y$. Before observing any data, we introduce a Gaussian distribution for $\boldsymbol\theta$:
\begin{equation}\label{eq:prior}
    \boldsymbol{\theta}\sim p(\boldsymbol{\theta})=\mathcal{N}(\boldsymbol{\theta}|\boldsymbol{\mu},\boldsymbol{\Sigma})\;.
\end{equation}
Bayesian linear regression~\cite{bayestheory_bishop2006pattern} assumes that the linear model has an approximation error $\epsilon$: $ y=\boldsymbol{\theta}^T\mathbf{x}+\epsilon$, and that the error $\epsilon$ follows a Gaussian distribution $\epsilon\sim\mathcal{N}(\epsilon|0,\sigma^2)$. For illustration purposes, we here assume that the values of $\{\boldsymbol{\mu},\boldsymbol{\Sigma},\sigma\}$ are provided. Next, considering that each data sample is conditionally independent on $\boldsymbol\theta$, we have the overall likelihood function~\cite{bayestheory_bishop2006pattern}:
\begin{equation}\label{eq:likelihood}
    p(\mathcal{D}|\boldsymbol{\theta})=\prod_{i=1}^N p(y_i\,|\,\boldsymbol{\theta},\mathbf{x}_i)=\prod_{i=1}^N\,\mathcal{N}(y_i|\boldsymbol{\theta}^T\mathbf{x}_i,\sigma^2)\;.
\end{equation}
Then according to Eq.~(\ref{eq:bayes_theorem2}), we can prove that the posterior $p(\boldsymbol{\theta}|\mathcal{D})$ also follows a Gaussian distribution~\cite{bayestheory_bishop2006pattern}:
\begin{equation}\label{eq:bayes_linreg_posterior}
    p(\boldsymbol{\theta}|\mathcal{D})=\mathcal{N}(\boldsymbol{\theta}|\boldsymbol{\mu}_\star,\boldsymbol{\Sigma}_\star)
\end{equation}
where
\begin{equation}\label{eq:expression_mu_star_bayes_linreg}
\begin{aligned}
    & \boldsymbol{\Sigma}_\star^{-1} = \boldsymbol{\Sigma}^{-1}+\frac{1}{\sigma^2}\mathbf{X}^T\mathbf{X}\\
    & \boldsymbol{\mu}_\star =  \boldsymbol{\Sigma}_\star(\boldsymbol{\Sigma}^{-1}\boldsymbol{\mu}+\frac{1}{\sigma^2}\mathbf{X}^T\mathbf{y})\;.\\ 
\end{aligned}
\end{equation}
Here $\mathbf{X}\in\mathbb{R}^{N\times H}$ is a matrix obtained by putting $\mathbf{x}_i$ at its $i$-th row, and $\mathbf{y}=[y_1,y_2,\cdots,y_N]^T\in\mathbb{R}^N$. Eq.~(\ref{eq:bayes_linreg_posterior}) depicts the posterior distribution of $\boldsymbol\theta$, calibrated based on the prior using the likelihood function. If we only want a point estimate of $\boldsymbol\theta$, then we can return the maximum-a-posteriori (MAP) estimator:
\begin{equation}\label{eq:map}
    \boldsymbol\theta_{\text{MAP}}=\argmax_{\boldsymbol\theta} p(\boldsymbol\theta |\mathcal{D})\;.
\end{equation}
Roughly speaking, $\boldsymbol\theta_{\text{MAP}}$ denotes the most probable value of $\boldsymbol\theta$ when we observe the dataset $\mathcal{D}$. In this case, since $p(\boldsymbol\theta |\mathcal{D})$ is Gaussian and the mode of a Gaussian distribution coincides with its mean, $\boldsymbol\theta_{\text{MAP}}=\boldsymbol\mu_\star$. Thus, our learned linear model is $f_{\boldsymbol{\theta}_{\text{MAP}}}(\mathbf{x})=\boldsymbol{\theta}_{\text{MAP}}^T\cdot\mathbf{x}$, which can be used to predict the value of $y$ at an unseen $\mathbf{x}$. Compared to the deterministic model, we can even obtain a probabilistic model if we want: we sample one instance of $\boldsymbol\theta$ from $p(\boldsymbol\theta|\mathcal{D})$ and then evaluate $\boldsymbol{\theta}^T\mathbf{x}$, and repeat $M$ times, obtaining $M$ different predictions for one single $\mathbf{x}$. Mathematically, this process is summarized by the predictive distribution:
\begin{equation}\label{eq:predictive}
    \mathbb{E}[f_{\boldsymbol{\theta}}(\mathbf{x})]=\int f_{\boldsymbol{\theta}}(\mathbf{x})\, p(\boldsymbol\theta|\mathcal{D}) \, d\boldsymbol\theta\;.
\end{equation}

Now let us consider several advanced concepts in Bayesian linear regression. To begin with, recall that we have not explained how to set $\{\boldsymbol{\mu},\boldsymbol\Sigma,\sigma\}$. In principle, the values of $\{\boldsymbol{\mu},\boldsymbol\Sigma\}$ reflect our belief about how $\boldsymbol\theta$ is distributed prior to observing any data, and $\sigma$ reflects how noisy the observed data are. Usually, $\{\boldsymbol{\mu},\boldsymbol\Sigma,\sigma\}$ are referred to as \emph{hyperparameters}, since their values largely impact the posterior $p(\boldsymbol{\theta}|\mathcal{D})$ as shown in Eq.~(\ref{eq:bayes_linreg_posterior}). In the Bayesian literature, there are two ways to determine their values: (i)~a standard Bayesian treatment: we directly set them according to some prior knowledge; and (ii)~an empirical Bayesian treatment: we estimate them from the dataset $\mathcal{D}$~\cite{bayestheory_bishop2006pattern,empiricalbayes_casella1985introduction,empiricalbayes_morris1983parametric} such as by using expectation-maximization (EM) algorithm~\cite{bayestheory_bishop2006pattern}. At first glance, empirical Bayes might sound deviant, as it sets the prior using the observed data. In fact, empirical Bayes raised debates among statisticians when it was proposed, but nowadays it is commonly agreed that empirical Bayes has its own merits. For instance, the James-Stein estimator~\cite{empiricalbayes_JS_stein1955inadmissibility}, derived using empirical Bayes, is widely accepted in high-dimensional statistics. 

Secondly, in cases when prior knowledge about the problem of concern is unknown, introducing a prior might be impossible or unwanted. Consequently, we can directly maximize the likelihood function without introducing a prior (or considering the posterior), which is known as maximum likelihood estimation (MLE):
\begin{equation}\label{eq:mle}
    \boldsymbol{\theta}_{\text{MLE}}=\argmax_{\boldsymbol\theta} p(\mathcal{D}|\boldsymbol\theta)\;.
\end{equation}
In the example of linear regression, using Eq.~(\ref{eq:likelihood}), we can show $\boldsymbol{\theta}_{\text{MLE}}=(\mathbf{X}^T\mathbf{X})^{-1}\mathbf{X}\mathbf{y}$, recovering an ordinary least square estimator~\cite{bayestheory_bishop2006pattern}. In general, it has been shown that MLE has a close connection to the cross entropy and mean squared error loss function used in deep learning (see Chapter 3.1.1 and 4.2.2 of~\cite{bayestheory_bishop2006pattern}).


Thirdly, in many applications as demonstrated by Eq.~(\ref{eq:predictive}), besides the MAP estimator, we also want to sample from the posterior distribution. This is trivial for the linear toy example since the posterior is known to be Gaussian. However, the posterior does not always have an analytical form in other applications.\footnote{It depends on the form of prior and likelihood. Given a likelihood function, it is possible to carefully design a prior, making the posterior analytical. Such a prior is called \emph{conjugate prior}~(see Chapter 2.4.2 of \cite{bayestheory_bishop2006pattern}).} When it does not, sampling from it becomes challenging. To tackle this issue, \emph{approximate inference} (e.g., variational inference, expectation propagation) and \emph{sampling methods} (e.g., importance sampling, MCMC) are usually applied. See, respectively, Chapters 10 and 11 of~\cite{bayestheory_bishop2006pattern} for more details about these methods. Additionally, it is important to note that while some of these methods are specific to the Bayesian framework, they can also be applied outside of it. For instance, sampling methods can sample from any distribution, not confined to a posterior. However, in a broad sense, these approaches also belong to Bayesian methods. As will be reviewed later, they are frequently used to resolve problems in the EDA domain.

\subsection{Gaussian Process Regression and Bayesian Optimization}

An improvement to the Bayesian linear regression presented in the previous subsection is to first pre-select a feature mapping function $\boldsymbol{\phi}(\mathbf{x})$, and next fit a linear mapping: $f_{\boldsymbol\theta}(\mathbf{x})=\boldsymbol{\theta}^T\boldsymbol{\phi}(\mathbf{x})$ in the feature space~\cite{bayestheory_bishop2006pattern} instead of in the original design space. In this way, Bayesian linear regression can learn a nonlinear relationship between $\mathbf{x}$ and $y$. However, how to wisely choose a feature mapping $\boldsymbol{\phi}(\cdot)$ is not trivial.\footnote{Using a neural network to solve a regression problem can also be comprehended as automatically learning a good feature mapping $\boldsymbol\phi(\cdot)$ (i.e., representation) from the training data.}

\begin{figure}
    \centering
    \includegraphics[width=1.0\linewidth]{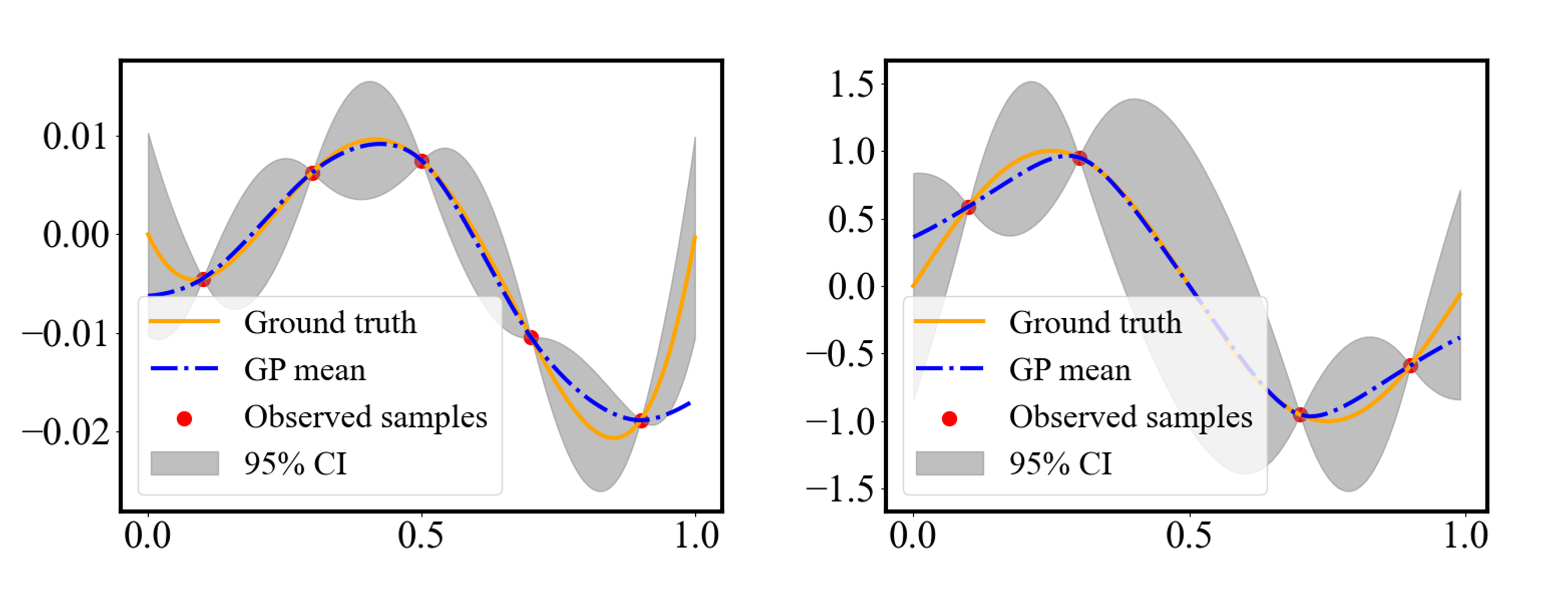}
    \caption{Use of a GPR model to fit a fourth-order polynomial with zeros at $\{0,0.2,0.6,1.0\}$ (left) and a sinusoidal function (right). The gray area is the 95\% confidence interval $[\mu_\star-1.96\sigma_\star,\mu_\star+1.96\sigma_\star]$.}
    \label{fig:gp_demo}
\end{figure}

Gaussian process regression (GPR) is one methodology implicitly choosing a feature mapping $\boldsymbol\phi(\cdot)$. In what follows, we will adopt the Bayesian view to formulate a GPR model. As in the previous subsection, suppose a dataset $\mathcal{D}=\{(\mathbf{x}_i, y_i)|i=1,2,\cdots, N\}$ is given. To define a GPR model, we need to specify a scalar mean function $m(\mathbf{x})$ and a scalar kernel function $k(\mathbf{x},\mathbf{x}^\prime)$. GPR assumes any finite number of target values follow a Gaussian distribution:
\begin{equation}
    \mathbf{y}=[y_1,y_2,\cdots,y_N]^T\sim p(\mathbf{y})=\mathcal{N}(\mathbf{y}|\boldsymbol{\mu},\mathbf{K})\;,
\end{equation}
where $\boldsymbol\mu\in \mathbb{R}^N$ is the mean vector, and its $i$-th entry equals $m(\mathbf{x}_i)$.
Here $\mathbf{K}\in\mathbb{R}^{N\times N}$ is the covariance matrix, and the entry on its $i$-th row and $j$-th column equals $k(\mathbf{x}_i,\mathbf{x}_j)$. If a new $\mathbf{x}_\star$ is given, then the $(N+1)$-dimensional vector made up of its corresponding target value $y_\star$ and the previous $\mathbf{y}$ also follows a Gaussian distribution due to the GPR assumption. Then, based on the product rule of probability, GPR provides a probabilistic prediction for the objective value $y_\star$ of a new input $\mathbf{x}_\star$:
\begin{equation}\label{eq:gpr_expression}
\begin{aligned}
    \mathcal{GP}(\mathbf{x}_\star)=p(y_\star(\mathbf{x}_\star)\,|\,y_1,\cdots,y_N)\sim \mathcal{N}(\mu_\star,\sigma_\star^2)\;.
\end{aligned}
\end{equation}
where ${\mu}_\star\in\mathbb{R}$ and ${\sigma}_\star^2\in\mathbb{R}^{+}$ are both functions of $\mathbf{x}_\star$~\cite{gaussianprocess_williams2006gaussian}:
\begin{equation}\label{eq:def_mu_std_in_gpr}
\left\{
\begin{aligned}
    {\mu}_\star(\mathbf{x}_\star) &=m(\mathbf{x}_\star) + k(\mathbf{x}_\star,\mathbf{X})\cdot\mathbf{K}^{-1}\cdot(\mathbf{y}-\boldsymbol\mu)\\
    \sigma_\star^2(\mathbf{x}_\star)&=k(\mathbf{x}_\star,\mathbf{x}_\star) -k(\mathbf{x}_\star,\mathbf{X})\cdot\mathbf{K}^{-1}\cdot k(\mathbf{X},\mathbf{x}_\star)\\
\end{aligned}\right.
\end{equation}
where $k(\mathbf{x}_\star,\mathbf{X})$ is a row vector in $\mathbb{R}^N$:
\begin{equation}
k(\mathbf{x}_\star,\mathbf{X})=\left[k(\mathbf{x}_\star,\mathbf{x}_1),k(\mathbf{x}_\star,\mathbf{x}_2)\cdots,k(\mathbf{x}_\star,\mathbf{x}_N)\right]
\end{equation}
and $k(\mathbf{X},\mathbf{x}_\star)=k(\mathbf{x}_\star,\mathbf{X})^T$ is a column vector in $\mathbb{R}^N$~\cite{bo_siliconphotonics_gao2022automatic}. In short, once $m(\mathbf{x})$ and $k(\mathbf{x},\mathbf{x}^\prime)$ are specified and the dataset $\mathcal{D}$ is given, GPR provides a probabilistic prediction for the objective value $y_\star$ of a new input $\mathbf{x}_\star$ following Eq.~(\ref{eq:gpr_expression})-(\ref{eq:def_mu_std_in_gpr}). A common choice is a zero mean function $m(\mathbf{x})=0$ and a radial basis kernel $k(\mathbf{x},\mathbf{x}^\prime)=\exp{(-a{||\mathbf{x}-\mathbf{x}^\prime||^2})}$, where $a$ is a hyperparameter, which can be estimated following the aforementioned empirical Bayesian treatment~\cite{gaussianprocess_williams2006gaussian}. Note that in a GPR model, although we have not explicitly defined a feature mapping $\boldsymbol\phi(\cdot)$, the kernel function $k(\mathbf{x},\mathbf{x}^\prime)$ implicitly defines one, subject to the relation: $k(\mathbf{x},\mathbf{x}^\prime)=\boldsymbol{\phi}(\mathbf{x})^T\boldsymbol{\phi}(\mathbf{x}^\prime)$. For example, the radial basis kernel corresponds to a feature mapping with an infinite feature dimension (see Chapter 6.2 of~\cite{bayestheory_bishop2006pattern} for details). For other technical details of GPR, refer to~\cite{gaussianprocess_schulz2018tutorial,gaussianprocess_williams2006gaussian,gaussianprocess_neal2012bayesian,bayestheory_bishop2006pattern,bo_shahriari2015taking}. Fig.~\ref{fig:gp_demo} demonstrates how we use a GPR model to fit a fourth-order polynomial and a sinusoidal function.

Using GPR as a surrogate model, Bayesian optimization (BO) is an efficient global optimization technique, especially suitable for an expensive-to-evaluate objective function~\cite{bo_frazier2018tutorial,bo_shahriari2015taking,bo_origin_movckus1975bayesian}. Specifically, consider the following optimization problem:\footnote{To maintain consistency in our review, all optimization problems will be treated as minimization problems, because a maximization problem can be addressed by minimizing the objective function with its sign flipped.}
\begin{equation}
    \min_{\mathbf{x}\in\Gamma}\, y=l(\mathbf{x})
\end{equation}
where $\Gamma\subseteq\mathbb{R}^H$ represents the feasible design space, and the function $l(\cdot)$ is time-consuming to invoke. Algo.~\ref{algo:bayesopt} shows the major steps of Bayesian optimization, rephrased from~\cite{bo_siliconphotonics_gao2022automatic}. Note that the function $l(\cdot)$ is needed only in steps 1 and 5, while in step 4, the acquisition function only needs the GPR model. Many choices of acquisition functions are available in the literature, among which lower evidence (or confidence) bound (LCB) might be the simplest:
\begin{equation}
    \text{LCB}(\mathbf{x})=\mu_\star(\mathbf{x})-\gamma\cdot\sigma_\star(\mathbf{x})
\end{equation}
where $\gamma\in\mathbb{R}^{+}$ is a user-predefined constant, balancing the importance of the two terms. When minimizing LCB, we attempt to find a new design $\mathbf{x}_{\text{new}}$ at two kinds of regions: where $\mu_\star(\mathbf{x})$ is small, or where $\sigma_\star(\mathbf{x})$ is large. The former corresponds to~\emph{exploitation} --- finding the global minimum of $l(\mathbf{x})$ based on the current GPR model, while the later corresponds to~\emph{exploration} --- refining the GPR model by sampling in regions where it is uncertain~\cite{bo_frazier2018tutorial,bo_shahriari2015taking,bo_siliconphotonics_gao2022automatic}. It is important to emphasize that the surrogate model and the acquisition function are the two crucial components of BO, and there are numerous implementation variations of both in the existing literature. For instance, a Bayesian neural network (BNN) can serve as the surrogate model. Besides LCB, examples of acquisition functions include expected improvement (EI) and probability of improvement (PI), among others.

\begin{figure}[!thb]
    \centering
    \includegraphics[width=1.0\linewidth]{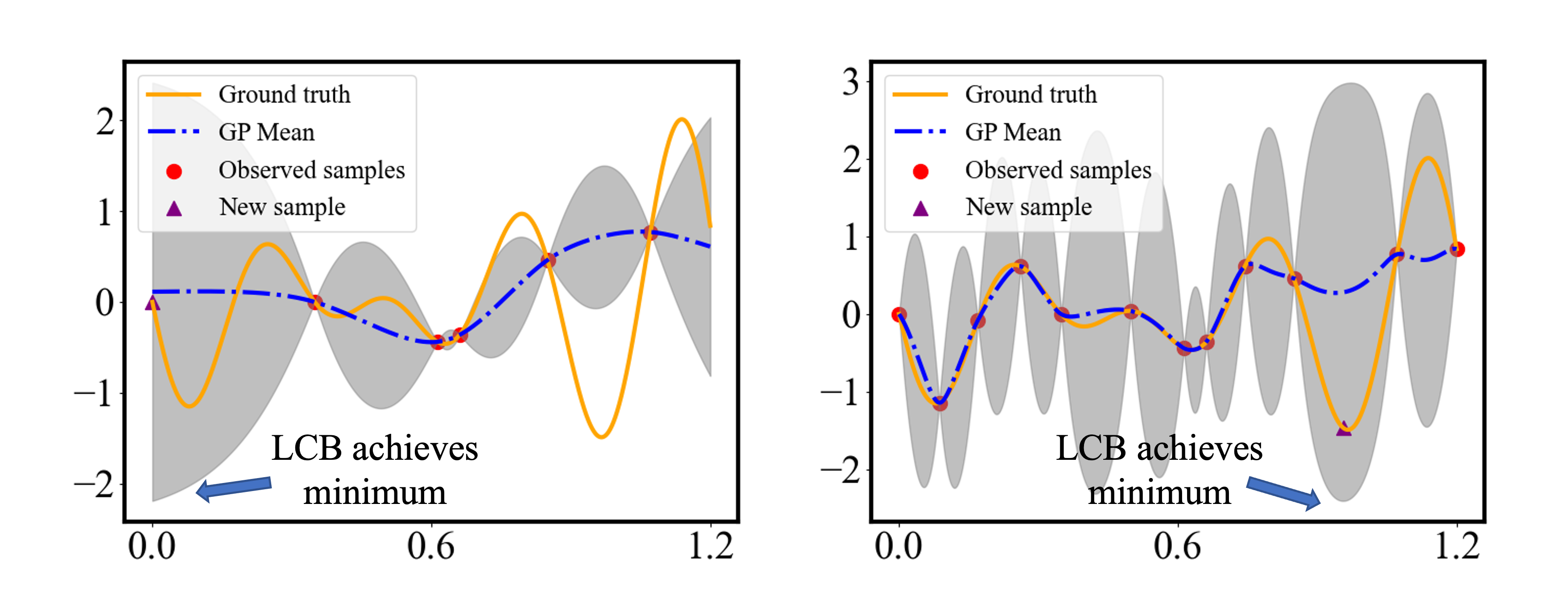}
    \caption{Use of Bayesian optimization to minimize a one-dimensional toy function. We set $N_{\text{init}}=5$ and $\gamma=5$. The shaded gray area represents $[\mu_{\star}-\gamma\sigma_\star,\mu_{\star}+\gamma\sigma_\star]$ in this figure. Left: After randomly sampling $5$ initial samples, we begin the Bayesian optimization loop. Right: Bayesian optimization yields a good design at the 13-th sample.}
    \label{fig:my_label}
\end{figure}

\begin{algorithm}[!thb]
\caption{Bayesian Optimization}\label{algo:bayesopt}
\begin{algorithmic}[1] 
\STATE Specify the initial and the maximum sample $N_{\text{init}}$ and $N_{\text{max}}$, respectively.
\STATE Sample $N_{\text{init}}$ designs from $\Gamma$ and evaluate their objective values, yielding $\mathcal{D}=\{(\mathbf{x}_i,y_i=l(\mathbf{x}_i))|i=1,2,\cdots,N_{\text{init}}\}$.
\FOR{$N_{\text{cur}}=N_{\text{init}}+1:1:N_{\text{max}}$}
\STATE Build a Gaussian process regression model $\mathcal{GP}(\mathbf{x})$ based on $\mathcal{D}$ using Eq.~(\ref{eq:gpr_expression})-(\ref{eq:def_mu_std_in_gpr}).
\STATE Minimize an acquisition function with respect to $\mathbf{x}$, yielding a new design $\mathbf{x}_{\text{new}}$.
\STATE Evaluate $y_{\text{new}}=l(\mathbf{x}_{\text{new}})$ and add $(\mathbf{x}_{\text{new}},y_{\text{new}}))$ into $\mathcal{D}$.
\ENDFOR
\STATE Return the design with the smallest function value in $\mathcal{D}$.
\end{algorithmic}
\end{algorithm}

\begin{table}[!htb]
    \centering
    \caption{A summary of important concepts related to the Bayesian framework}
    \begin{tabular}{c|c}
    \toprule
    Concept & Meaning or Use \\
    \midrule
    prior, likelihood, posterior & Eq.~(\ref{eq:bayes_theorem2}) \\
    MLE & Eq.~(\ref{eq:mle})\\
    MAP & Eq.~(\ref{eq:map}) \\
    Conjugate prior & The $p(\boldsymbol{\theta})$ s.t. $p(\boldsymbol{\theta}|\mathcal{D})$ is analytical \\
    Empirical Bayes & Set hyper-parameters using $\mathcal{D}$\\
    Sampling methods & Sample from non-analytical $p(\boldsymbol{\theta}|\mathcal{D})$ \\
    Approximate inference & Approximate non-analytical $p(\boldsymbol{\theta}|\mathcal{D})$ \\
    Gaussian process regression (GPR) & Eq.~(\ref{eq:gpr_expression})\\
    Bayesian optimization (BO) & Algo.~\ref{algo:bayesopt} \\
    \cdashline{1-2}[1pt/1pt]
    Expectation maximization (EM) & Find local MLE or MAP estimator \\
    K-means algorithm & Partition the dataset into $K$ clusters \\
    Bayesian neural network (BNN) & A probabilistic regression model \\
    Belief network & A probabilistic graphical model \\
    \bottomrule
    \end{tabular}
    \label{tab:summary_bayes_concept}
\end{table}

Table~\ref{tab:summary_bayes_concept} provides a summary of key concepts related to the Bayesian framework that will be frequently used in our review, and the abbreviations there will be assumed to be familiar to the reader. In the following, we briefly explain the concepts below the dashed dividing line, which have not been explained in our recap. 

Expectation maximization (EM) is an iterative process to calculate the MLE or MAP estimator, where the model of interest has unknown \emph{latent variables} (i.e., random variables associated with the model but not observed). It can be regarded as a local optimization method, and sometimes it is also used to estimate hyperparameters. 

The K-means algorithm can be applied to partition the provided data into $K$ disjoint groups. It is identical to applying the EM method to a Gaussian mixture (see Chapter 9 of~\cite{bayestheory_bishop2006pattern}). 

A Bayesian neural network (BNN) extends a modern neural network with Bayesian inference. Specifically, it introduces a distribution on the neural network weight, and consequently, the training of a neural network turns into performing posterior inference. It is identical to the function of a GPR model in the sense that they both can provide a probabilistic prediction for a given sample. 

A belief network (also known as a Bayes net) represents the conditional independencies of a set of random variables using a directed acyclic graph. It is useful for representing complex, uncertain systems, and can be used for various tasks, such as inferring conditional probabilities. 

For a more in-depth understanding of the concepts in Table~~\ref{tab:summary_bayes_concept}, we recommend consulting the comprehensive text~\cite{bayestheory_bishop2006pattern}. Finally, it is important to note that some of the methods listed in Table~\ref{tab:summary_bayes_concept} are not exclusive to the Bayesian framework, such as sampling methods and approximate inference.

\section{Preliminary}\label{sec:preliminary}

Next, we formally introduce a few mathematical notations, so that all papers that will be reviewed can be explained under the same mathematical framework. We then provide some high-level remarks on why Bayesian methods were bound to
be introduced to EDA in the recent several decades.

\subsection{A Unified Mathematical Framework for Review}

To begin, we use $\mathbf{y}\in\mathbb{R}^{H_y}$ to represent the circuit performances of interest (e.g., gain, phase margin, or slew rate of an Op-Amp). Without loss of generality, 
we can collect all variables impacting $\mathbf{y}$ into a vector $\mathbf{z}\in\mathbb{R}^{H}$, and define the mapping from $\mathbf{z}$ to $\mathbf{y}$ by a function $\mathbf{f}(\cdot):\mathbb{R}^{H}\to\mathbb{R}^{H_y}$, i.e., $\mathbf{y}=\mathbf{f}(\mathbf{z})$.\footnote{Note that in our review, we usually treat $\mathbf{z}$ as a stationary variable that does not change over time. However, including time dependence $\mathbf{z}=\mathbf{z}(t)$ is necessary for certain problems (such as estimating time-to-failure in reliability analysis).} As an example, $\mathbf{z}$ can represent circuit component connection, supply voltage, design parameters (e.g., transistor length/width, variables of parametric cells), and working temperature. In a real-world scenario, the function $\mathbf{f}$ is unknown, but fortunately, contemporary circuit simulators are sufficiently accurate to work as an approximation. In most cases, we focus on one single circuit performance (i.e., $H_y=1$), and when it happens, we will respectively use $y$ and $f$ to substitute $\mathbf{y}$ and $\mathbf{f}$. 

Without considering any stochasticity or when the randomness is small enough to be negligible, the function $\mathbf{f}$ and the variable $\mathbf{z}$ are both deterministic. However, this is not the case in modern IC design, and now it is usually assumed that $\mathbf{f}$ is deterministic while $\mathbf{z}$ is stochastic. The randomness of $\mathbf{z}$ is rooted in the variations introduced during manufacturing, such as line edge roughness, pattern density effect, and variation of transistor width/length or oxide thickness~\cite{dfm_boning1999models,dfm_agarwal2007characterizing}. Mathematically, for ease of later description, we can adopt a reparameterization, decomposing $\mathbf{z}$ into two parts:
\begin{equation}\label{eq:reparam}
    \mathbf{z}=\mathbf{x}+\boldsymbol{\epsilon}
\end{equation}
where $\mathbf{x}\in\mathbb{R}^H$ is a deterministic variable, and $\boldsymbol{\epsilon}\in\mathbb{R}^H$ is a stochastic variable following a probability density function $p(\boldsymbol{\epsilon})$. Two clarifications must be made regarding the validity of the above reparameterization. First, any random variable can be written in this way when doing a mean shift,\footnote{As an example, a random variable $w$ follows a Gaussian distribution $\mathcal{N}(1.5,4)$ can be written as $w=1.5+\epsilon$, where $\epsilon\sim\mathcal{N}(0,4)$.} so the above reparameterization is mathematically valid. Second, Eq.~(\ref{eq:reparam}) also has physical meanings. Here $\mathbf{x}$ can be comprehended as design variables (e.g., transistor width) whose values we can specify in circuit design, while $\boldsymbol{\epsilon}$ represents the amount of process variation added during manufacturing. Note that this reparametrization has a strong representation power, covering a few corner cases. For instance, as we mentioned, entries of $\mathbf{z}$ might represent quantities that are not impacted by process variations (e.g., the circuit component connection) and hence are deterministic. To handle this case, the corresponding entries in $\boldsymbol{\epsilon}$ can be set to a Dirac function.

\begin{figure}[!htb]
    \centering
    \includegraphics[width=0.9\linewidth]{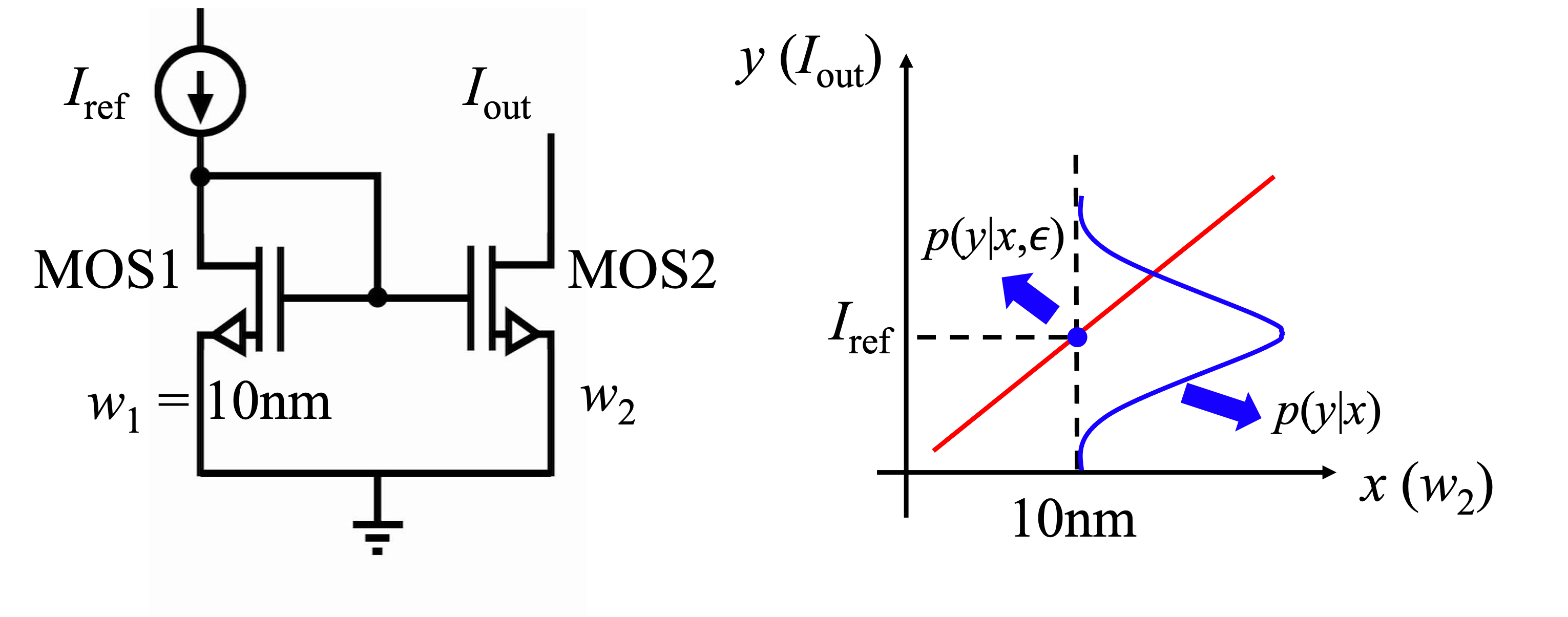}
    \caption{A current mirror example. Assume that the width of MOS2 is the design variable $x$ and the output current is the performance of interest $y$, and that a Gaussian distributed variation $\epsilon$ is added to $x$. {Left}: A simplified schematic of a current mirror. The output current is given by $I_{\text{out}}=w_2\cdot I_{\text{ref}}/w_1$. {Right}: The red line represents the function $f(x)=(I_{\text{ref}}/w_1) x$. The blue solid dot represents $p(y|x=10,\epsilon=0)=\delta(y-f(10))$, implying that $y=I_{\text{ref}}$ given $x=10$ and $\epsilon=0$. The blue Gaussian curve represents the distribution $p(y|x=10)$.}
    \label{fig:framework_demo}
\end{figure}

The distribution $p(\boldsymbol{\epsilon})$ can be regarded as given in a process design kit (PDK) file. Now, we describe several quantities of interest, and more details will be covered later in our review. First, given a specific $\mathbf{x}$ and an $\boldsymbol{\epsilon}$ instance sampled from $p(\boldsymbol{\epsilon})$, we know $\mathbf{y}=\mathbf{f}(\mathbf{z})=\mathbf{f}(\mathbf{x}+\boldsymbol{\epsilon})$. This fact can be described with the help of Dirac function:
\begin{equation}\label{eq:p_of_y_conditioned_x_epsilon}
    p(\mathbf{y}|\mathbf{x},\boldsymbol{\epsilon})=\delta(\mathbf{y}-\mathbf{f}(\mathbf{x}+\boldsymbol{\epsilon}))\;.
\end{equation}
Namely, it indicates that conditioned on $\mathbf{x}$ and $\boldsymbol{\epsilon}$, $\mathbf{y}$ is deterministic and attains $\mathbf{f}(\mathbf{x}+\boldsymbol{\epsilon})$. We emphasize that since $\boldsymbol{\epsilon}$ is in the conditioning, Eq.~(\ref{eq:p_of_y_conditioned_x_epsilon}) implicitly assumes $\boldsymbol{\epsilon}$ is already fixed and thus $\mathbf{y}$ is deterministic. See Fig.~\ref{fig:framework_demo} for an illustrating example. Now, if we wish to take the randomness of $\boldsymbol{\epsilon}$ into consideration, we should make $\mathbf{y}$ only conditioned on $\mathbf{x}$. Namely, the quantity of interest now is $p(\mathbf{y}|\mathbf{x})$. Using the product rules of conditional probability, we have:
\begin{equation}\label{eq:p_of_y_conditioned_x}
\begin{aligned}
    p(\mathbf{y}|\mathbf{x}) &= \int p(\mathbf{y},\boldsymbol{\epsilon}|\mathbf{x})\,d\boldsymbol{\epsilon}=\int p(\mathbf{y}|\mathbf{x},\boldsymbol{\epsilon})\,p(\boldsymbol{\epsilon})\,d\boldsymbol{\epsilon}\\
    &=\int \delta(\mathbf{y}-\mathbf{f}(\mathbf{x}+\boldsymbol{\epsilon}))\, p(\boldsymbol{\epsilon})\,d\boldsymbol{\epsilon}\;.\\
\end{aligned}
\end{equation}
This conditional probability reveals the fact that, even though we fix the value of $\mathbf{x}$, the circuit performance $\mathbf{y}$ is still probabilistic due to the randomness of $\boldsymbol{\epsilon}$. Refer to Fig.~\ref{fig:framework_demo} for an intuitive understanding of Eq.~(\ref{eq:p_of_y_conditioned_x_epsilon}) and (\ref{eq:p_of_y_conditioned_x}). 

Estimating $p(\mathbf{y}|\mathbf{x})$ is one key question in circuit performance modeling. Once it is known, it can be further exploited to calculate several important metrics. As two examples, first, the expectation of $p(\mathbf{y}|\mathbf{x})$ is of interest, since it indicates the probable values of $\mathbf{y}$:\footnote{Chebyshev's inequality guarantees that the observed value of a random variable is closed to its expectation with a high probability.}

\begin{equation}
\begin{aligned}
    \mathbb{E}[p(\mathbf{y}|\mathbf{x})]&=\int\int\delta(\mathbf{y}-\mathbf{f}(\mathbf{x}+\boldsymbol{\epsilon}))\, p(\boldsymbol{\epsilon})\,d\boldsymbol{\epsilon}\,d\mathbf{y}\\
    &=\int \mathbf{f}(\mathbf{x}+\boldsymbol{\epsilon})p(\boldsymbol{\epsilon})\, d\boldsymbol{\epsilon} = \mathbb{E}[\mathbf{f}(\mathbf{x}+\boldsymbol{\epsilon})]
\end{aligned}
\end{equation}
where $\mathbb{E}[\cdot]$ denotes the expectation operator. Note that strictly, the expectations on the first line and the last line are taken with respect to $\mathbf{y}$ and $\boldsymbol{\epsilon}$, respectively. Second, when a circuit performance specification is known, the parametric yield $P_s$ of a circuit can be calculated using $p(\mathbf{y}|\mathbf{x})$:
\begin{equation}\label{eq:yield_expression}
    {P}_{s}=\int_{\mathbf{y}\in{\Omega}} p(\mathbf{y}|\mathbf{x}) \,d\mathbf{y}
\end{equation}
where $\Omega\subset\mathbb{R}^{H_y}$ represents the region where $\mathbf{y}$ satisfies the performance specification. Usually, $\Omega$ is a hyper-cube represented by $\Omega=\{\mathbf{y}\in\mathbb{R}^{H_y}|\mathbf{L}\leq\mathbf{y}\leq\mathbf{U}\}$, where $\mathbf{L}\in\mathbb{R}^{H_y}$ and $\mathbf{U}\in\mathbb{R}^{H_y}$ denote the lower bound and upper bound of the acceptable circuit performance value. Later, we will show how various problems are defined within this framework.


\subsection{High-Level Remarks on Bayesian Methods in EDA}

Before going into detailed review of the literature, here we provide some high-level remarks on why Bayesian methods were bound to be introduced to EDA over the recent several decades. In our understanding, three main reasons motivate adoption: 
\begin{itemize}\itemsep3pt
    \item \textbf{Stochastic process variation}: The impact of stochastic process variation becomes non-negligible, and rigorous analysis is desired. Dealing with randomness requires a statistical framework, for which Bayesian methods are well-suited.
    \item\textbf{High sample cost}: We need to obtain several data samples to solve circuit modeling or optimization problems, but data acquisition is time-consuming or economically expensive. Fortunately, provided a proper prior, Bayesian methods can achieve good results with only a few samples.
    \item \textbf{Accessible prior information}: Due to the hierarchical (e.g., schematic then layout) and correlated (e.g., multiple process corners) characteristics of circuit design, a prior distribution is relatively easy to define and works well in practice.

\end{itemize}

Let us start with the first motivation above. As summarized in Moore's law, the process node of MOSFET technology has kept scaling down ever since its birth. Around 2003, the process node, for the first time, reached below 100~nm. This led to denser integration, but process variation also became a larger concern~\cite{dfm_boning1999models,dfm_borkar2005designing,dfm_sarangi2008varius,dfm_Michael2008design,dfm_kuhn2011process}. For example, the authors in~\cite{dfm_boning1999models} show that the within-die variation of the transistor's effective channel length increases from 25\% in 1997 to around 50\% in 2007. The author~\cite{dfm_borkar2005designing} argues that the threshold voltage of transistors will distribute more widely in an advanced technology node.

As  clear from the previous subsection, the consideration of stochastic process variation $\boldsymbol{\epsilon}$ makes almost all quantities of interest in a circuit problem probability-related. For example, we summarize the expressions of circuit performance $y$ under the case of either considering process variation or not in the following equation:
 \begin{equation}
 \begin{aligned}
    &\text{w/o variation: }&y= f(\mathbf{z})=&f(\mathbf{x})\quad &\boldsymbol{\epsilon}=\mathbf{0} \\
     &\text{w/ variation: }&p(\mathbf{y}|\mathbf{x})& &\boldsymbol{\epsilon}\sim p(\boldsymbol{\epsilon})\;.\\
 \end{aligned}
 \end{equation}
It reveals that we need tools to handle quantities related to distributions. For this goal, Bayesian methods (in its broad sense) is an excellent candidate framework for these stochastic problems, as they are a set of approaches focusing on inferring unknown parameters from observations and learning or sampling from distributions.




Related to the second motivation summarized earlier, in both circuit modeling and optimization, we need to acquire data to solve the problem of concern. However, the challenge lies in that data acquisition (i.e., calling the function $\mathbf{f}$) is achieved by circuit simulation or real chip measurement, which is often time-consuming and economically expensive. Consequently, we want an approach that can resolve circuit-related problems with as few samples as possible. This motivates the usage of Bayesian methods. For instance, Bayesian optimization is usually deployed in a gradient-free manner, and it is commonly believed that it can achieve better optimized objective value with fewer samples compared to genetic algorithms, particle swarm optimization, or simulated annealing~\cite{bocircuitopt_lyu2018multi,bo_frazier2018tutorial,bo_siliconphotonics_gao2022automatic}. 

As another illustrative example, we show how a prior can help in the scenario of few data. Consider the problem of modeling the probability of an uneven coin landing on its head. Denote this probability as $\mu$, and assume we know $\mu\in[0.4,0.6]$ due to the coin material and weight distribution, prior to any toss. The conventional formula to estimate this probability is $\mu=m/N$, where $m$ is the number of the coin head coming up and $N$ is the total number of experiments. In a few-data scenario, we toss the coin twice (i.e., $N=2$) and suppose that we observe the coin lands on head in both experiments (i.e., $m=2$). Then, we will wrongly estimate $\mu=m/N=1$. However, with a Bayesian framework, we can introduce the prior belief using a Beta distribution parameterized by $a=b=5$, i.e., $\text{Beta}(\mu|a,b)$, corresponding to our belief $\mu=0.5$, which we choose based on the middle of $[0.4,0.6]$. Then, the MAP estimator is given by $\mu_{\text{MAP}}=(m+a)/(m+a+l+b)=0.58$~\cite{bayestheory_bishop2006pattern}, which seems more reasonable.\footnote{Note that as long as $a=b$, it corresponds to the prior belief $\mu=0.5$. Here the particular number $a=b=5$ is only for illustration purposes. Note that the values of $a$ and $b$, termed the effective number of observations, balances the weight of the prior and observed data in the MAP estimator. See Chapter 2.1 of~\cite{bayestheory_bishop2006pattern} for details.}

Regarding the third motivation for adoption of Bayesian methods in EDA, it is notable that the current IC design flow exhibits hierarchical and correlated characteristics, which facilitate the definition of a prior. For instance, when designing a circuit, it is common practice to commence with schematic-level simulation. Upon satisfying all specifications, it is necessary to extract the parasitic effects and perform post-layout simulation, which provides a more accurate representation of the performance of the fabricated circuit, albeit at a longer simulation duration. Therefore, we can utilize the data obtained from schematic-level simulation to establish a prior, thereby aiding in the analysis during the post-layout phase. 

In addition to the aforementioned time-hierarchical property, correlated characteristic are also present in modern IC design. Specifically, to ensure circuit functionality, it is necessary to simulate a circuit at various process corners. The performance of the circuit at these different corners exhibits a close correlation - if an Op-Amp circuit achieves a gain of 35~dB at a certain corner, it is likely that the gain at the other corners will also be around 35~dB. By encoding this information into a prior distribution, we can effectively leverage it and speed up multi-corner circuit modeling~\cite{yield_multicorner_gao2019efficient}.

\section{Modeling, Analysis, and Diagnosis}\label{sec:modeling}

In this section, we demonstrate how to utilize Bayesian methods to address circuit modeling, analysis, and diagnosis problems. The papers reviewed in this section are classified into several themes, which are noted to be overlapping and non-exhaustive. The categorization of each paper is based on the primary keywords specified in the original paper.

\subsection{Performance Modeling}

Three particular circuit performance modeling problems have received substantial attention: (i)~given a design $\mathbf{x}_\star$, estimate $p(y|\mathbf{x}_\star)$ (or the moments) as a function of $y$; (ii)~estimate $f(\mathbf{x}_\star+\boldsymbol{\epsilon})$ as a function of $\boldsymbol{\epsilon}$ when $\mathbf{x}_\star$ is given; and (iii)~estimate $\mathbb{E}[p(y|\mathbf{x})]$ as a function of $\mathbf{x}$. Table~\ref{tab:summary_performance_modeling} summarizes papers on Bayesian circuit performance modeling according to the performed task. For Case I, a trivial solution is to sample $\{\boldsymbol{\epsilon}_i|i=1,2,\cdots,N\}$ and call the function $f(\cdot)$ at each $(\mathbf{x}_\star+\boldsymbol{\epsilon}_i)$, giving $\{y_i=f(\mathbf{x}_\star+\boldsymbol{\epsilon}_i)|i=1,2,\cdots,N\}$. Then, we can exploit any probability distribution fitting technique to learn $p(y|\mathbf{x}_\star)$. A similar approach can be adopted for other cases. Nevertheless, the bottleneck for such approaches lies in the fact that evaluating $f(\cdot)$ is often time-consuming or economically expensive. Namely, we might only be able to afford to call $f(\cdot)$ a few times~\cite{perfmodel_firstbmf_li2013bayesian}, leading to a poor fit.

Bayesian methods have been introduced to address this pain point. The majority of these works are motivated by the fact that evaluating $f(\cdot)$ might be expensive, but its surrogate $f^\mathcal{L}(\cdot)$, which is cheap to evaluate yet similar to $f(\cdot)$, is usually accessible. From now on, when such a surrogate is available, we will use a superscript $\mathcal{L}$, short for ``low fidelity,'' to denote the quantities related to it. Correspondingly, we will use $f^\mathcal{H}(\cdot)$ to substitute the original notation $f(\cdot)$ for consistency, where $\mathcal{H}$ is short for ``high fidelity.'' Table~\ref{tab:lowhigh_fidelity} summarizes a few low/high-fidelity cases in circuits side-by-side. Now, taking Case I as an example, the basic idea of a Bayesian approach is to first estimate the distribution $p(y^\mathcal{L}|\mathbf{x}_\star)$ with a dataset $\{y_i^\mathcal{L}=f^\mathcal{L}(\mathbf{x}_\star+\boldsymbol{\epsilon}_i)|i=1,2,\cdots, N^\mathcal{L}\}$. Because evaluating $f^\mathcal{L}(\cdot)$ is cheap, $N^\mathcal{L}$ can be large and $p(y^\mathcal{L}|\mathbf{x}_\star)$ can be well-fit. Next, we can encode $p(y^\mathcal{L}|\mathbf{x}_\star)$ into a prior distribution and learn the posterior, which corresponds to the desired $p(y^\mathcal{H}|\mathbf{x}_\star)$, with only a few samples $\{y_i^\mathcal{H}=f^\mathcal{H}(\mathbf{x}_\star+\boldsymbol{\epsilon}_i)|i=1,2,\cdots, N^\mathcal{H}\}$ (where $N^\mathcal{H}<<N^\mathcal{L}$). In the following, we review a few representative papers in detail.

\begin{table}[!htb]
    \centering
    \caption{A summary of papers on Bayesian circuit performance modeling}
    \begin{tabular}{c|c}
    \toprule
    Task & References \\
    \midrule
    Case I: estimate $p(y|\mathbf{x}_\star)$ & \cite{perfmodel_firstbmf_li2013bayesian,perfmodel_moment_gu2013efficient,perfmodel_huang2015efficient,perfmodel_mp_li2014efficient,perfmodel_tao2019large,perfmodel_zhang2010toward,perfmodel_bmf_li2012efficient} \\
    Case II: estimate $f(\mathbf{x}_\star+\boldsymbol{\epsilon})$     & \cite{perfmodel_bmf_wang2013bayesian,perfmodel_dualpriorbmf_huang2016efficient,perfmodel_cbmf_wang2016correlated,perfmodel_cbmf_gao2022correlated, perfmodel_clbmf_wang2015co}\\
    Case III: estimate $\mathbb{E}[p(y|\mathbf{x})]$   & \cite{perfmodel_selfhealing_sun2013indirect,perfmodel_liu2020transfer} \\
    \bottomrule
    \end{tabular}
    \label{tab:summary_performance_modeling}
\end{table}

\begin{table}[!htb]
    \centering
    \caption{A comparison of low/high-fidelity cases in circuits}
    \begin{tabular}{c|c}
    \toprule
    Low Fidelity  $f^\mathcal{L}(\cdot)$       & High Fidelity $f^\mathcal{H}(\cdot)$ \\
    \midrule
    Loose stopping criterion    & Strict stopping criterion \\
    W/o parasitic effect (schematic-level) & W/ parasitic effect (post-layout)\\
    Circuit simulation           & Lab measurement \\
    An outdated technology node & An advanced technology node \\
    \bottomrule
    \end{tabular}\label{tab:lowhigh_fidelity}
\end{table}

The idea of using Bayesian inference to facilitate performance modeling of ICs was first proposed in~\cite{perfmodel_bmf_wang2013bayesian,perfmodel_firstbmf_li2013bayesian} around 2013, under the term Bayesian model fusion (BMF)~\cite{perfmodel_bmf_wang2013bayesian,perfmodel_firstbmf_li2013bayesian}. For instance, the authors in~\cite{perfmodel_firstbmf_li2013bayesian} consider the problem of estimating the mean of $p(y^\mathcal{H}|\mathbf{x}_\star)$. They assume that both $p(y^\mathcal{H}|\mathbf{x}_\star)$ and $p(y^\mathcal{L}|\mathbf{x}_\star)$ follow Gaussian distributions: $p(y^\mathcal{H}|\mathbf{x}_\star)=\mathcal{N}(\mu^\mathcal{H},(\sigma^\mathcal{H})^2)$ and $p(y^\mathcal{L}|\mathbf{x}_\star)=\mathcal{N}(\mu^\mathcal{L},(\sigma^\mathcal{L})^2)$, so the question asks the value of $\mu^\mathcal{H}$. In their context, high fidelity $\mathcal{H}$ refers to post-layout simulation, while low fidelity $\mathcal{L}$ refers to schematic-level simulation. Following the previous discussion, since there is little computational cost when sampling low-fidelity data, the mean $\mu^\mathcal{L}$ and the standard deviation $\sigma^\mathcal{L}$ can be accurately estimated. Then, the authors encode the knowledge that $\mu^\mathcal{H}$ should not be far from $\mu^\mathcal{L}$ using a prior distribution: $p(\mu^\mathcal{H})=\mathcal{N}(\mu^\mathcal{L},\sigma_0^2)$, where $\sigma_0$ is an unknown hyper-parameter estimated by cross validation in~\cite{perfmodel_firstbmf_li2013bayesian}. They sample a few high-fidelity samples $\mathcal{D}^\mathcal{H}=\{y_1^\mathcal{H},y_2^\mathcal{H},\cdots,y_{N^\mathcal{H}}^\mathcal{H}\}$, based on which a likelihood function $p(\mathcal{D}^\mathcal{H}|\mu^\mathcal{H})$ can be defined. Next, using Bayes' theorem, they calculate the posterior $p(\mu^\mathcal{H}|\mathcal{D}^\mathcal{H})\propto p(\mathcal{D}^\mathcal{H}|\mu^\mathcal{H})p(\mu^\mathcal{H})$, which further gives the MAP estimator of $\mu^\mathcal{H}$. They show that this MAP estimator achieves a $2.25\times$ sampling cost reduction or $1.5\times$ error reduction, compared with a direct calculation. The authors' Gaussian assumption on $p(y^\mathcal{H}|\mathbf{x}_\star)$ and $p(\mu^\mathcal{H})$ guarantees that the posterior is analytically tractable and the MAP estimator has a closed form. However, we warn the readers that careful considerations must be taken when assuming so, since $p(y^\mathcal{H}|\mathbf{x}_\star)$ does not always distribute symmetrically, let alone follow a Gaussian distribution. In~\cite{perfmodel_firstbmf_li2013bayesian}, they also consider the problem of estimating $p(y^\mathcal{H}|\mathbf{x}_\star)$ without the Gaussian restriction. They approach this by decomposing $p(y^\mathcal{H}|\mathbf{x}_\star)$ and $p(y^\mathcal{L}|\mathbf{x}_\star)$ using the same set of basis functions so that the coefficients ahead of the basis share a similar relationship as $\mu^\mathcal{L}$ and $\mu^\mathcal{H}$ do, and hence, the BMF framework can be applied. They further demonstrate that compared with kernel density estimation relying merely on $\mathcal{D}^\mathcal{H}$, about $10\times$ sampling cost is saved~\cite{perfmodel_firstbmf_li2013bayesian}.

The paper~\cite{perfmodel_bmf_wang2013bayesian} deals with estimating $f^\mathcal{H}(\mathbf{x}_\star+\boldsymbol{\epsilon})$ as a function of $\boldsymbol{\epsilon}$. Similar to~\cite{perfmodel_firstbmf_li2013bayesian}, it uses a decomposition: $f^\mathcal{U}(\mathbf{x}_\star+\boldsymbol{\epsilon})=\sum_{m=1}^M \alpha_m^\mathcal{U} g_m(\boldsymbol{\epsilon})$, where $\mathcal{U}$ can denote either $\mathcal{H}$ or $\mathcal{L}$, and $\{g_m(\boldsymbol{\epsilon})\,|\,m=1,2,\cdots,M\}$ are a set of pre-selected basis functions. Because $\boldsymbol{\epsilon}$ is usually high-dimensional~\cite{perfmodel_bmf_wang2013bayesian}, the authors use a zero-mean Gaussian prior $p(\alpha_m^\mathcal{H})=\mathcal{N}(0,(\alpha^\mathcal{L}_m)^2)$ to impose sparsity. Then, they sample a dataset $\mathcal{D}^\mathcal{H}=\{(\boldsymbol{\epsilon}_n,y^\mathcal{H}_n)|n=1,2,\cdots, N^\mathcal{H})\}$, define a likelihood function, calculate the posterior, and estimate the MAP estimator of all $\alpha_m^\mathcal{H}$s. Two implementation details regarding missing prior knowledge and estimating hyper-parameter using cross-validation are discussed. They demonstrate the effectiveness of the proposed method on a ring oscillator and an SRAM circuit with $\boldsymbol{\epsilon}$ possessing $7177$ and $6617$ dimensions, respectively. Compared with orthogonal mapping pursuit, excavating information from low-fidelity data with the proposed method attains $9\times$ run-time speedup~\cite{perfmodel_bmf_wang2013bayesian}.

Later in~\cite{perfmodel_selfhealing_sun2013indirect}, the authors consider the problem of indirect performance sensing, where they attempt to build a function mapping from several performances of measurement (PoM) to a performance of interest (PoI). They assume that an old model $f^\mathcal{L}(\cdot)$ before the process shift occurs exists. With a few new data sampled after the process shift occurs, a new model $f^\mathcal{H}(\cdot)$ can be calibrated from the old model $f^\mathcal{L}(\cdot)$ using Bayes' theorem. The authors demonstrate that incorporating this technique into an on-chip self-healing flow  can improve the yield of a VCO from 0\% to 69.17\%, and it requires fewer samples to reach a similar yield compared to other methods such as ordinary least square~\cite{perfmodel_selfhealing_sun2013indirect}.

The paper~\cite{perfmodel_dualpriorbmf_huang2016efficient} extends the BMF framework by incorporating an additional source of prior and names the proposed approach Dual-prior BMF (DP-BMF). Namely, they fuse two different prior models, and next combine and re-calibrate them
with a small number of high-fidelity training samples~\cite{perfmodel_dualpriorbmf_huang2016efficient}. A belief network is used to describe the proposed model, and cross-validation is used to determine the hyper-parameters controlling the importance of the two prior sources. It is shown that compared to a single prior, DP-BMF achieves around $1.83\times$ sampling cost reduction~\cite{perfmodel_dualpriorbmf_huang2016efficient}.

The authors in~\cite{perfmodel_huang2015efficient} consider several performance metrics at the same time. Mathematically, they attempt to model the moments of $p(\mathbf{y}^\mathcal{H}|\mathbf{x}_\star)$. Note that now $\mathbf{y}^\mathcal{H}$ is a vector representing several performance metrics, instead of a scalar representing one performance. They exploit a Gaussian likelihood function and a normal-Wishart prior distribution, and a MAP estimator is calculated. They demonstrate their method on an Op-Amp and an ADC. In both examples, five performance metrics are considered (i.e., $\mathbf{y}^\mathcal{H}\in\mathbb{R}^5$), and it is shown that around $16\times$ sampling cost reduction can be achieved, compared to a traditional maximum likelihood estimation method without using low-fidelity data.

The aforementioned works exploit low-fidelity information to assist with a task associated with high-fidelity. Notice that high fidelity is the predominant state that we care about, while low fidelity only provides auxiliary information. In addition to the works above, a few Bayesian circuit modeling works do not rely on the correlation of low-fidelity $\mathcal{L}$ and high-fidelity $\mathcal{H}$, or there might be more states than two, and each state is equally important --- quantities of interest exist in all states. For instance, Wang et al.~\cite{perfmodel_cbmf_wang2016correlated} consider the problem of learning $f(\mathbf{x}_\star+\boldsymbol{\epsilon})$. However, the challenge lies in that instead of one single $f$, they are actually concerned with learning $K$ similar function mappings $\{f^{1},f^{2},\cdots,f^{K}\}$ at the same time. In their context, $K$ represents the number of states (i.e., knob configurations) for a tunable analog/RF circuit (e.g., a down-conversion
mixer). Their key idea is that since these $K$ function mappings are correlated, a multivariate joint Gaussian prior distribution can be assumed to capture this information. This Gaussian prior implicitly correlates the modeling of $f^i$ with $f^j$ ($i\neq j$), potentially boosting the modeling accuracy. A MAP estimator is calculated to estimate the unknown model coefficients, and an expectation-maximization algorithm is used to determine the values of hyper-parameters~\cite{perfmodel_cbmf_wang2016correlated}. About $2\times$ sampling cost reduction is attained compared with the state-of-the-art orthogonal matching pursuit method~\cite{perfmodel_cbmf_wang2016correlated}. Later in~\cite{perfmodel_cbmf_gao2022correlated}, this method is further extended to do multi-corner performance modeling for analog/RF circuits, where $K$ represents the number of process-voltage-temperature (PVT) corners.

A similar idea has been applied in~\cite{perfmodel_moment_gu2013efficient} to estimate the first and second moments of $p(y|\mathbf{x}_\star)$ when the sample size is extremely small. It exploits the fact that data collected at different design stages, different configurations, and different corners are not independent, but are correlated~\cite{perfmodel_moment_gu2013efficient}. It first attempts to learn the prior following an Empirical Bayes' view, and then calculates the moments for each population. Compared to sample mean and standard deviation estimators, the proposed method achieves about $2\times$ error reduction when being tested on commercial I/O links~\cite{perfmodel_moment_gu2013efficient}.

Co-Learning BMF (CL-BMF)~\cite{perfmodel_clbmf_wang2015co} is proposed to take advantage of the coefficient side information (CSI) as well as performance side information (PSI) to do efficient circuit performance modeling. A Bayesian graphical model is constructed, where the CSI and PSI are encoded by the prior distribution and the likelihood function, respectively. It is shown that CL-BMF achieves $5\times$ cost reduction, compared to the original BMF method on a low noise amplifier example.

\subsection{Parametric Yield and Failure Rate Estimation}

Parametric yield estimation~\cite{yield_bmfbd_fang2014bmf,yield_rejimon2008probabilistic,yield_multicorner_gao2019efficient,yield_multicornerconjugate_shi2020multi} is closely related to circuit performance modeling as shown in Eq.~(\ref{eq:yield_expression}). Note that $P_s$ actually is a function of $\mathbf{x}$, i.e., $P_s=P_s(\mathbf{x})$ in Eq.~(\ref{eq:yield_expression}). For convenience, the concept of failure rate is also sometimes used:
\begin{equation}\label{eq:failure_expression}  {P}_{f}=P_f(\mathbf{x})=\int_{\mathbf{y}\notin{\Omega}} p(\mathbf{y}|\mathbf{x}) \,d\mathbf{y}\;.
\end{equation}
Comparing with Eq.~(\ref{eq:yield_expression}), we obviously have $0\leq P_s(\mathbf{x}), P_f(\mathbf{x})\leq 1$ and $P_s(\mathbf{x})+P_f(\mathbf{x})=1$. In a yield/failure rate estimation problem, we are usually concerned with the value of $P_s$ or $P_f$ at a specific design $\mathbf{x}=\mathbf{x}_\star$. Namely, we only desire a scalar value $P_s(\mathbf{x}_\star)$ or $P_f(\mathbf{x}_\star)$ here. Alternatively, learning the function $P_s(\mathbf{x})$ or $P_f(\mathbf{x})$ is needed in the subsequent yield/failure rate optimization subsection in Section~\ref{sec:optimization}. In this subsection, when there is no confusion, we will use the notation $P_{s,\star}$ and $P_{f,\star}$ as shorthand for $P_{s}(\mathbf{x}_\star)$ and $P_{f}(\mathbf{x}_\star)$, respectively. Note that in the literature, due to nomenclature differences, the use of the words `yield' and `failure rate' sometimes blend. In our review, we will strictly follow the definition that we give in Eqs.~(\ref{eq:yield_expression}) and (\ref{eq:failure_expression}).\footnote{Thus, papers such as~\cite{failure_sis_katayama2010sequential} are classified into failure rate estimation, though their original keywords are yield estimation.}

The trivial approach to estimating $P_{s,\star}$ is Monte Carlo sampling. Namely, we sample $\{\boldsymbol{\epsilon}_i|i=1,2,\cdots,N\}$ and call the function $f(\cdot)$ at each $(\mathbf{x}_\star+\boldsymbol{\epsilon}_i)$, yielding $\{y_i=f(\mathbf{x}_\star+\boldsymbol{\epsilon}_i)|i=1,2,\cdots,N\}$. Then, if there are $M$ samples $y_i$ located in the acceptable region $\Omega$, then we estimate $P_{s,\star}=M/N$. This Monte Carlo approach suffers from a sampling cost issue: the function $f(\cdot)$ is often expensive to run, and a large $N$ is not affordable. The first notable line of work follows the same thought as in the previous circuit performance modeling subsection. Namely, a cheap surrogate $f^\mathcal{L}(\cdot)$ of $f(\cdot)$ is leveraged, and the original $f(\cdot)$ will be denoted as $f^\mathcal{H}(\cdot)$ for consistency. Low-fidelity data is first used to estimate $P_{s,\star}^\mathcal{L}$. Because low-fidelity data is cheap to obtain, $P_{s,\star}^\mathcal{L}$ can be well-estimated. Then the information $P_{s,\star}^\mathcal{H}\approx P_{s,\star}^\mathcal{L}$ is encoded using a prior distribution, and the next estimate $P_{s,\star}^\mathcal{H}$ is found using a few high-fidelity data samples via the MAP estimator.

Let us give a few concrete examples. In~\cite{yield_bmfbd_fang2014bmf}, considering the natural of performance testing is binary (e.g., $\mathbf{y}$ is either in or not in $\Omega$), the authors set the likelihood of observing one sample $\mathbf{y}^\mathcal{H}$ (i.e., $p(\mathbf{y}^\mathcal{H}|P_{s,\star}^\mathcal{H})$) as a Bernoulli distribution. Then, the overall likelihood function is $p(\mathcal{D}^\mathcal{H}|P_{s,\star}^\mathcal{H})=\prod_{i=1}^{N^\mathcal{H}} p(\mathbf{y}_i^\mathcal{H}|P_{s,\star}^\mathcal{H})$, where $\mathcal{D}^\mathcal{H}=\{\mathbf{y}_i^\mathcal{H}\,|\,i=1,2,\cdots,N\}$ is the training dataset. Next, they introduce a Beta prior $p(P_{s,\star}^\mathcal{H})=p(P_{s,\star}^\mathcal{H}|a,b)$, where $\{a,b\}$ are hyper-parameters determined by $P_s^\mathcal{L}$ via optimization. Using a Beta distribution as the prior has the advantage of easy computation, because the Beta distribution is the conjugate prior for the Bernoulli distribution, and the conjugacy ensures that the posterior has an analytical form~\cite{bayestheory_bishop2006pattern}. Finally, the posterior is calculated via Bayes' theorem $p(P_{s,\star}^\mathcal{H}|\mathcal{D}_\mathcal{H})\propto p(\mathcal{D}_\mathcal{H}|P_{s,\star}^\mathcal{H})p(P_{s,\star}^\mathcal{H})$. Maximum-a-posteriori is performed and the MAP estimator of $P_{s,\star}^\mathcal{H}$ is returned. This proposed method, termed BMF-BD, is shown to achieve around $8\times$ sampling cost reduction compared to the trivial Monte Carlo estimate~\cite{yield_bmfbd_fang2014bmf}.

The authors in~\cite{yield_bigr_gao2021bayesian} notice that the low-fidelity and high-fidelity information share general features, and also possess their own unique features (see Fig.~1 in~\cite{yield_bigr_gao2021bayesian}). Thus, they design a customized graphical model containing one general region and two private regions to capture this observation. A regularized likelihood objective is maximized by a conditional expectation-maximization algorithm to infer all model parameters. It is shown that this proposed BI-GR method can achieve superior accuracy compared to Monte Carlo or BMF-BD~\cite{yield_bmfbd_fang2014bmf}.

In~\cite{yield_multicorner_gao2019efficient} and~\cite{yield_multicornerconjugate_shi2020multi}, the authors consider the multi-corner yield estimation problem. Namely, a set of yield values $\{P_{s,\star}^1,P_{s,\star}^2,\cdots,P_{s,\star}^K\}$ are desired to be estimated, where $K$ represents the number of process corners. Both of these two works are motivated by the fact that the values of $P_{s,\star}^i$ and $P_{s,\star}^j$ are correlated, so modeling all $K$ yields simultaneously can implicitly exploit this correlation and reduce sampling cost. They differ in what prior is used. Ref.~\cite{yield_multicorner_gao2019efficient} uses a multivariate Gaussian prior so that the covariance matrix naturally has the physical meaning of correlation. Concretely, the entry on the $i$-th column and $j$-th row of the covariance matrix represents the correlation between the $i$-th corner and $j$-th corner or that between $P_{s,\star}^i$ and $P_{s,\star}^j$. However, this Gaussian prior makes the posterior intractable, and a Laplacian approximation is adopted to do posterior inference~\cite{yield_multicorner_gao2019efficient}. Alternatively, \cite{yield_multicornerconjugate_shi2020multi} exploits a conjugate prior, so the posterior has a closed form and the computation is straightforward. Both these works demonstrate better accuracy or around $2\times$ reduced sampling cost when compared with Monte Carlo.

\begin{table}[!htb]
    \centering
    \caption{A summary of papers on yield or rare failure rate estimation}
    \begin{threeparttable}
    \begin{tabular}{c|c|c}
    \toprule
    &  Yield & Rare Failure Rate \\
    \midrule
      \thead{w/ P-L-P \\ (MAP or MLE)}   &  \cite{yield_bmfbd_fang2014bmf,yield_multicorner_gao2019efficient,yield_multicornerconjugate_shi2020multi,yield_bigr_gao2021bayesian} &  \cite{failure_zhai2018efficient,failure_bsss_sun2015fast,failure_sss_sun2015fast,failure_multicorner_gao2019efficient,failure_tsss_gao2022fast,failure_posterior_parijat2014leveraging,failure_bsus1_bect2017bayesian} \\
    \midrule
\thead{w/o P-L-P \\ (Sampling method)}  &  $\times$ &  \cite{failure_weller2019bayesian,failure_is_qazi2010loop,failure_hefenbrock2020fast,failure_weller2019bayesian1,failure_rescope_wu2014rescope,failure_wu2014fast,failure_statisticalblockade_singhee2009statistical,failure_sis_katayama2010sequential,failure_gibbs_dong2011efficient,failure_fonseca2010statistical,failure_abu2008methodology,failure_kanj2006mixture,failure_bofailure_Hu2018parallelizable,failure_testselection_parijat2016using,failure_bo_hu2019enabling,failure_gong2012fast,failure_wu2016hyperspherical,failure_is_shi2018fast,failure_shi2019adaptive,failure_is_shi2020non,failure_sus_sun2014fast,failure_sus2_sun2015fast,failure_sus3_au2001estimation,failure_ape_tao2017correlated,failure_apa_yu2016efficient} \\
    \bottomrule
    \end{tabular}
    \label{tab:yield_summary}
    \begin{tablenotes}
    \footnotesize
    \item $^*$ P-L-P is short for prior-likelihood-posterior. `w/ P-L-P' implies that the Bayesian framework (i.e., maximum-a-posteriori) or at least likelihood function and maximizing likelihood estimation is used. 
    \end{tablenotes}
    \end{threeparttable}
\end{table}

In the works reviewed above, estimating $P_{s,\star}$ or $P_{f,\star}$ is usually interchangeable and makes no difference. However, we emphasize that there is a certain case when dealing with $P_{f,\star}$ is more important than $P_{s,\star}$. That is when $P_{s,\star} \to 1$ or equivalently $P_{f,\star}\to 0$ (e.g., $P_{f,\star}=10^{-8}$), and the corresponding problem is termed rare failure rate estimation~\cite{failure_bsss_sun2015fast,failure_is_qazi2010loop,failure_sss_sun2015fast}. Such a case often occurs in an SRAM circuit or other large-scale circuits containing repeated components~\cite{failure_bsss_sun2015fast,failure_is_qazi2010loop,failure_sss_sun2015fast}. This rare failure rate estimation problem has received substantial research attention over the past 20 years, and various methods (e.g., sampling methods~\cite{failure_sus3_au2001estimation,failure_sss_sun2015fast,failure_is_qazi2010loop}, classification~\cite{failure_statisticalblockade2_singhee2008recursive,failure_statisticalblockade_singhee2009statistical,failure_rescope_wu2014rescope}, meta-modeling~\cite{failure_metamodel_shi2019metamodel,failure_metamodel_jian2015efficient}, and analytical models~\cite{failure_analytical_mukhopadhyay2005modeling,failure_analytical_james2013leveraging,failure_analytical_weckx2014nmc}) have been adopted for it. Considering the scope of our review, we will focus mainly on sampling methods, and leave other types of methods for readers to explore themselves. Note that some of the sampling methods might not use the Bayesian prior-likelihood-posterior framework, but since sampling methods belong to Bayesian methods in a broad sense, we feel the necessity to also include and review them here. Table~\ref{tab:yield_summary} summarizes papers on yield or rare failure rate estimation according to if a prior-likelihood-posterior framework or a sampling method is used or dominates the approach. In the following, we will detail several classical rare failure rate estimation methods.

For later description convenience, we re-write the quantity of interest in the following equation:
\begin{equation}\label{eq:rarefailure_expression}
\begin{aligned}
P_{f,\star}&=P[\mathbf{y}\notin\Omega]=\int I(\mathbf{y})p(\mathbf{y}|\mathbf{x}_\star)\,d\mathbf{y}\;,\\
\end{aligned}
\end{equation}
where $P[\cdot]$ represents the probability of the event occurring, and $I(\mathbf{y})$ is an indicator function, which equals $1$ when $\mathbf{y}\notin\Omega$, and $0$ otherwise. We emphasize that there is another equivalent form for Eq.~(\ref{eq:rarefailure_expression}):
\begin{equation}\label{eq:rarefailure_expression2}
\begin{aligned}
P_{f,\star}&=P[\mathbf{f}(\mathbf{x}_\star+\boldsymbol{\epsilon})\notin\Omega]=\int I(\boldsymbol{\epsilon}) p(\boldsymbol{\epsilon})\,d\boldsymbol{\epsilon}\;,\\
\end{aligned}
\end{equation}
where $\boldsymbol{\epsilon}$ is the stochastic variation defined in Eq.~(\ref{eq:reparam}). Here $I(\boldsymbol{\epsilon})$ is still an indicator function, but its domain is in the space of $\boldsymbol{\epsilon}$.\footnote{We have not introduced a subscript or a superscript to distinguish between  $I(\boldsymbol{\epsilon})$ and $I(\mathbf{y})$ in Eqs.~(\ref{eq:rarefailure_expression})-(\ref{eq:rarefailure_expression2}), but readers should notice that they are not the same function. $I(\boldsymbol{\epsilon})$ and $I(\mathbf{y})$ are associated with the acceptable region in the space of $\boldsymbol\epsilon$ and $\mathbf{y}$, respectively.} Namely, $I(\boldsymbol{\epsilon})$ equals $1$ if $f(\mathbf{x}_\star+\boldsymbol{\epsilon})\notin\Omega$, and $0$ otherwise. Thus, Eq.~(\ref{eq:rarefailure_expression}) defines $P_{f,\star}$ in the space of $\mathbf{y}$, while Eq.~(\ref{eq:rarefailure_expression2}) defines $P_{f,\star}$ in the space of $\boldsymbol{\epsilon}$. They are equivalent and related by a change of variable using Eq.~(\ref{eq:p_of_y_conditioned_x_epsilon}). For consistency with the published literature, we will use the Eq.~(\ref{eq:rarefailure_expression2}) framing from now on.

The Monte Carlo approach estimates the integral in Eq.~(\ref{eq:rarefailure_expression2}) by drawing $N$ i.i.d. samples from $p(\boldsymbol{\epsilon})$ and calculating:
\begin{equation}\label{eq:rarefailure_mc}
        P_{f,\star}\approx \frac{1}{N}\sum_{i=1}^N I(\boldsymbol{\epsilon}_i),\quad \boldsymbol{\epsilon}_i\sim p(\boldsymbol{\epsilon})\;.
\end{equation}
Its main drawback is the large sampling cost. For instance, if $P_{f,\star}=10^{-6}$, which is common in a rare failure rate estimation problem, it roughly says one of $10^6$ $\boldsymbol{\epsilon}_i$ will fail the specification test (i.e., $I(\boldsymbol{\epsilon}_i)=1$). This implies that $N$ should be at least larger than $10^6$. Otherwise, it is highly likely that we wrongly observe that all samples pass the specification test (i.e., $I(\boldsymbol{\epsilon}_i)=0$) and trivially estimate $P_{f,\star}=0$.

\begin{figure}[!htb]
    \centering
    \includegraphics[width=1.0\linewidth]{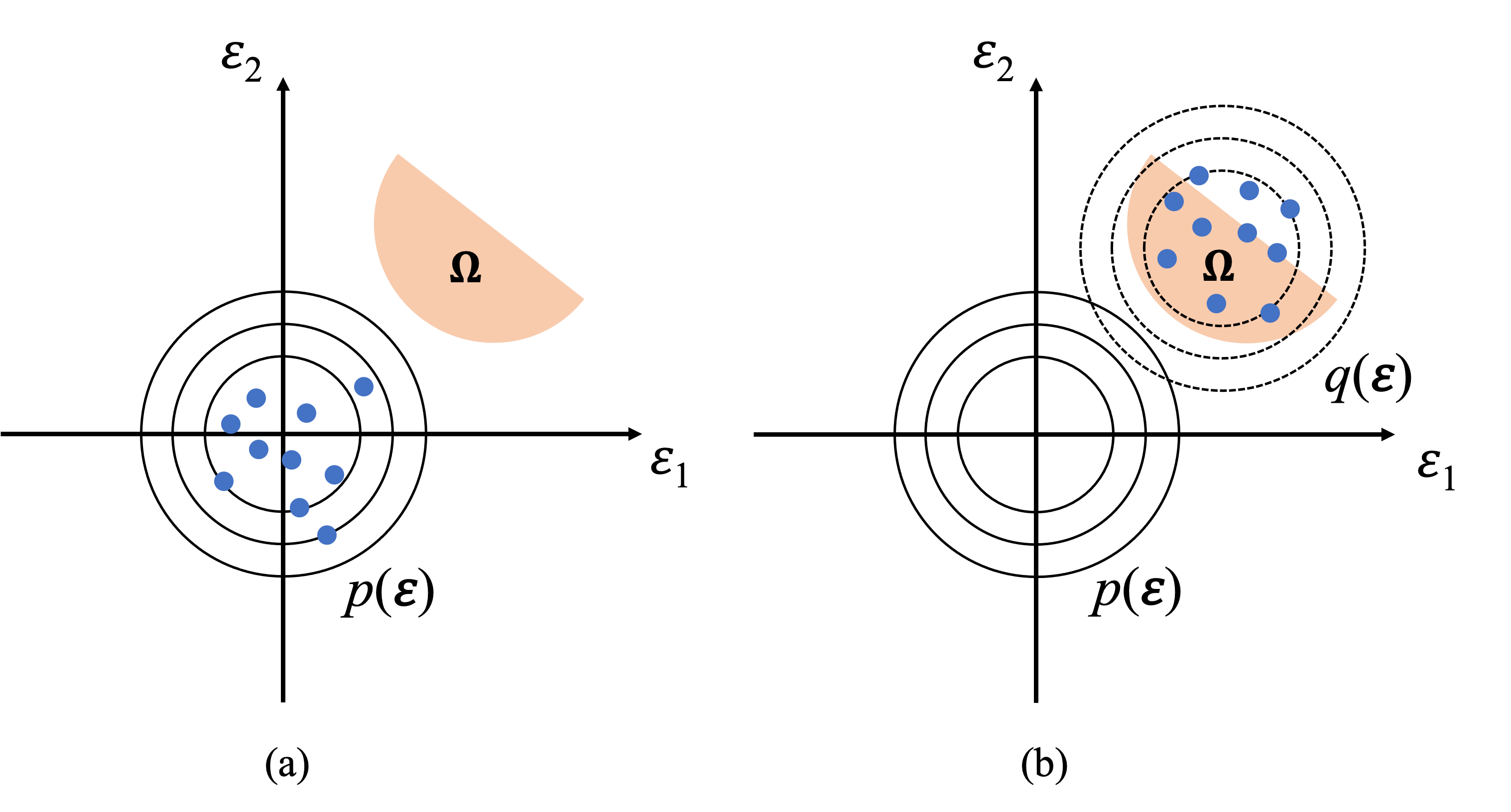}
    \caption{A demonstration of rare failure rate estimation in the two-dimensional Euclidean space $\boldsymbol{\epsilon}=[\epsilon_1,\epsilon_2]^T$, where $p(\boldsymbol{\epsilon})$ is a Gaussian distribution centered at the origin. $\Omega$ locates far from the origin, and $q(\boldsymbol{\epsilon})$ is assumed to be a Gaussian distribution centered around $\Omega$. Left: Monte Carlo approach. Right: Importance sampling.}
    \label{fig:rarefailure_is}
\end{figure}

Importance sampling~\cite{failure_is_shi2020non,failure_is_shi2018fast,failure_kanj2006mixture,failure_hefenbrock2020fast,failure_weller2019bayesian1,failure_weller2019bayesian,failure_is_qazi2010loop} might be the most frequently used rare failure event sampling technique. 
It requires a well-designed proposal distribution $q(\boldsymbol{\epsilon})$. Then, it draws $N$ i.i.d samples from the proposal $q(\boldsymbol{\epsilon})$ (instead of the original $p(\boldsymbol{\epsilon})$) and adopts a re-weighted formula:

\begin{equation}
     \begin{aligned}
    P_{f,\star}&=\int I(\boldsymbol{\epsilon}) p(\boldsymbol{\epsilon})\,d\boldsymbol{\epsilon}\\
    &=\int \left[I(\boldsymbol{\epsilon}) \frac{p(\boldsymbol{\epsilon})}{q(\boldsymbol{\epsilon})}\right]q(\boldsymbol{\epsilon})\,d\boldsymbol{\epsilon}\\
    &\approx \frac{1}{N}\sum_{i=1}^N I(\boldsymbol{\epsilon}_i)\frac{p(\boldsymbol{\epsilon}_i)}{q(\boldsymbol{\epsilon}_i)},\quad \boldsymbol{\epsilon}_i\sim q(\boldsymbol{\epsilon})\;.
     \end{aligned}
\end{equation}
The quantity ${p(\boldsymbol{\epsilon}_i)}/q(\boldsymbol{\epsilon}_i)$ is termed the importance weight, and it corrects the bias introduced by sampling from the proposal distribution~\cite{bayestheory_bishop2006pattern}. Fig.~\ref{fig:rarefailure_is} demonstrates rare failure rate estimation in the two-dimensional Euclidean space by Monte Carlo and the importance sampling method. In this example, the sample $\boldsymbol{\epsilon}_i$ drawn by Monte Carlo almost always has $I(\boldsymbol{\epsilon}_i)=0$. In contrast, with the proposal $q(\boldsymbol{\epsilon})$ shown in Fig.~\ref{fig:rarefailure_is}(b), importance sampling tends to obtain meaningful samples $\boldsymbol{\epsilon}_i$ with $I(\boldsymbol{\epsilon}_i)\neq 0$. It is obvious from the example that a good $q(\boldsymbol{\epsilon})$ should emphasize the region $\Omega$ so that more effective samples can be generated. However, the challenge here is we do not know where $\Omega$ is in a circuit problem, and thus setting a good $q(\boldsymbol{\epsilon})$ becomes a demanding question. 

To this end, many research efforts have searched for an answer. In~\cite{failure_is_qazi2010loop}, the authors apply mean shift to the distribution $p(\boldsymbol{\epsilon})$ and take the resulting distribution as the proposal. As an example, if $p(\boldsymbol{\epsilon})=\mathcal{N}(\mathbf{0},\mathbf{I})$, then they use $q(\boldsymbol{\epsilon})=\mathcal{N}(\mathbf{s},\mathbf{I})$, where $\mathbf{s}$ is a mean shift vector~\cite{failure_is_qazi2010loop} required to be determined. The value of $\mathbf{s}$ is chosen as the point on
the pass-fail boundary closest in quadratic distance to the
nominal operating point~\cite{failure_is_qazi2010loop}. In~\cite{failure_weller2019bayesian,failure_weller2019bayesian1}, the authors use a mixed Gaussian distribution, e.g., $q(\boldsymbol{\epsilon})=\sum_{i=1}^K a_i\mathcal{N}(\mathbf{s}_i,\sigma^2\mathbf{I})$, as the proposal. Then, they use Bayesian optimization to obtain a probabilistic model for the circuit performance and further utilize it to obtain the set of optimal mean shifts $\{\mathbf{s}_i\,|\,i=1,2,\cdots,K\}$. In~\cite{failure_kanj2006mixture}, the authors use a mixture distribution (not necessarily a mixture of Gaussians) as the proposal. Then they set the optimal mean shift using the calculated center of gravity of failures~\cite{failure_kanj2006mixture}.

In~\cite{failure_is_shi2018fast}, adaptive importance sampling is proposed. The authors still use a Gaussian mixture model $q(\boldsymbol{\epsilon})=\sum_{i=1}^K a_i\mathcal{N}(\mathbf{s}_i,\boldsymbol{\Sigma}_i^{-1})$ as the proposal. They iteratively update $\{\mathbf{s}_i\,|\,i=1,2,\cdots,K\}$ and $\{\boldsymbol{\Sigma}_i\,|\,i=1,2,\cdots,K\}$ using the newly drawn samples according to certain formulas. Later, the same authors~\cite{failure_is_shi2020non} further extend the method by considering a mixture of von Mises-Fisher distributions as the proposal in order to handle high-dimensional variation space. The unknown parameters of the proposal are solved by the expectation-maximization algorithm under the framework of maximum likelihood estimation~\cite{failure_is_shi2020non}. 

In~\cite{failure_gis_thomas2018gradient}, the authors propose gradient importance sampling. Specifically, they utilize the gradient to assist with finding the optimal mean shift, as the gradient reflects the sensitivity of the circuit performance with respect to the process variation. All of the aforementioned methods show superior accuracy compared to Monte Carlo in their papers. Readers can refer to the related work section of these papers for more work on importance sampling.

Subset simulation~\cite{failure_sus3_au2001estimation} is one of the most cited rare failure rate estimation approaches in the literature. It decomposes the calculation of $P_{f,\star}$ into estimating a series of conditional probabilities by gradually extending the boundary of $\Omega$. For illustration purposes, we consider one single performance $y$ and $\Omega=\{y\in \mathbb{R}\,|\,L < y < \infty\}$. After choosing a monotonically increasing sequence $\{L_i|i=0,1,\cdots,M\}$ satisfying $L_i\leq L_{i+1}$ and $L_0=L$, we notice the following important relation:
\begin{equation}
\begin{aligned}
    P[y\leq L_i]
    &=P[y\leq L_i,\,y\leq L_{i+1}]\\
    &=P[y\leq L_i\,|\,y\leq L_{i+1}]\cdot P[y \leq L_{i+1}]\;,\\
\end{aligned}
\end{equation}
where $P[\cdot]$ represents the event probability. The first line is because if $y\leq L_i$, then $y$ must be smaller than both $L_i$ and $L_{i+1}$, and the reverse statement also holds true. The second line uses the product rule of probability (see Eq.~(\ref{eq:original_bayes_theorem}) and the explanation below). Using this relation, we can derive:
\begin{equation}
\begin{aligned}
    P_{f,\star}
    &=P[y\notin \Omega]=P[y\leq L]=P[y\leq L_0]\\
    &=P[y\leq U_0|y\leq L_1]\cdot P[y \leq L_1]\\
    &=P[y\leq U_0|y\leq L_1]\, P[y \leq L_1 | y \leq L_2]\, P[y \leq L_2]\\
    &=P[y\leq L_M]\cdot \prod_{i=0}^{M-1} P[y\leq L_i|y\leq L_{i+1}]\;.
\end{aligned}
\end{equation}
Namely, subset simulation estimates $P[y\leq L_M]$ and $M$ conditional probabilities, and recovers the $P_{f,\star}$ using the equation above. It enjoys the benefit that the probabilities required to estimate are now relatively large, so estimating them is an easier task compared to directly estimating the extremely small $P_{f,\star}$~\cite{failure_sus_sun2014fast,failure_sus2_sun2015fast,failure_sus3_au2001estimation,failure_bsus1_bect2017bayesian}. For instance, to estimate $P_{f,\star}\approx 10^{-6}$, we can set $M=5$ and wisely choose $\{L_i|i=1,2,\cdots,M\}$ so that $P[y\leq L_M]\approx10^{-1}$ and $P[y\leq L_i|y\leq L_{i+1}]\approx 10^{-1}$. Estimating $P[y\leq L_M]\approx10^{-1}$ can be efficiently done by the Monte Carlo approach. Estimating the other conditional probabilities requires a sampling approach such as the Metropolis-Hastings algorithm~\cite{failure_sus_sun2014fast,failure_sus2_sun2015fast,failure_sus3_au2001estimation,failure_bsus1_bect2017bayesian}.

Scaled sigma sampling~\cite{failure_sss_sun2015fast,failure_bsss_sun2015fast,failure_tsss_gao2022fast} is another rare failure rate estimation approach designed for Gaussian or truncated Gaussian distributed variation. For illustration convenience, we here consider $p(\boldsymbol{\epsilon})=\mathcal{N}(\mathbf{0},\mathbf{I})$. These approaches scale up the standard deviation by a scaling factor $s$ and write the corresponding failure rate under the distorted distribution $p(\boldsymbol{\epsilon},s)=\mathcal{N}(\mathbf{0},s^2\mathbf{I})$:
\begin{equation}\label{eq:sss_scaled}
    P_{f,\star}(s)=\int I(\boldsymbol{\epsilon})p(\boldsymbol{\epsilon},s)\,d\boldsymbol{\epsilon}\;.
\end{equation}
They notice that a certain model template:
\begin{equation}\label{eq:sss_model}
    \ln P_{f,\star}(s)=\alpha+\beta\ln s +\frac{\gamma}{s^2}
\end{equation}
can be derived by approximating Eq.~(\ref{eq:sss_scaled}) via numerical differentiation. Here $\{\alpha,\beta,\gamma\}$ are unknown model parameters~\cite{failure_sss_sun2015fast}. Motivated by this finding, scaled sigma sampling first defines a set of scaling factors $\{s_i\,|\,i=1,2,\cdots,Q\}$ and then adopts Monte Carlo to simulate the corresponding $\{P_{f,\star}(s_i)\,|\,i=1,2,\cdots,Q\}$. Next, the unknown model parameters $\{\alpha,\beta,\gamma\}$ can be estimated using the set of data pair $\{(s_i,P_{f,\star}(s_i))|i=1,2,\cdots,Q\}$ by maximum likelihood estimation. Finally, we can substitute $s=1$ into Eq.~(\ref{eq:sss_model}) to get the $P_{f,\star}$ required to be estimated. The advantage of this approach lies in the fact that $P_{f,\star}(s_i)$ is relatively large (compared to $P_{f,\star}$) and easy to estimate by the Monte Carlo method.

In addition to the aforementioned works, a few other notable works also exist in the literature. In~\cite{failure_apa_yu2016efficient,failure_ape_tao2017correlated}, the authors relate the overall failure rate of an SRAM circuit (i.e., the target $P_{f,\star}$ in their context) with the failure rate of each SRAM cell or several cells, referred to as partial failure rate in~\cite{failure_apa_yu2016efficient}. APA is proposed in~\cite{failure_apa_yu2016efficient} to treat this relation as a linear function, while APE~\cite{failure_ape_tao2017correlated} does a further derivation and shows that the relationship is nonlinear. Both APA and APE extend the subset simulation to calculate partial failure rates of different orders. Next, APA calculates the overall SRAM circuit failure rate $P_{f,\star}$ by a linear weighted summation of these partial failure rates, while APE uses a nonlinear least square with a lookup table to solve $P_{f,\star}$. In~\cite{yield_multicorner_gao2019efficient}, the authors utilize the Bayesian framework together with scaled sigma sampling and the relation found by APA to estimate several failure rates at multiple process corners simultaneously. In~\cite{failure_gibbs_dong2011efficient}, Gibbs sampling is used to estimate the rare failure rate. In~\cite{failure_testselection_parijat2016using}, the authors develop a test-set selection method to best use pre-silicon knowledge and an
information-theory-based parameter ranking scheme to maximize the probability of observing post-silicon failures. In~\cite{failure_bofailure_Hu2018parallelizable}, the authors propose to use Bayesian optimization equipped with a novel acquisition function to effectively retrieve the worst-case rare performance. Later, the same authors further extend the method to deal with high-dimensional variation space in~\cite{failure_bo_hu2019enabling}.

To end this section, we emphasize that the rare failure rate estimation problem arises in many domains (e.g., reliability engineering, disaster/weather prediction), and approaches such as weighted ensemble, line sampling, and forward flux sampling are proposed there. We have only covered the major methods used in EDA. Readers can search for additional relevant literature using the keyword ``rare event sampling.''

\subsection{Spatial Variation Analysis}

The objective of spatial variation analysis is to identify and quantify the variability of parameters within a single wafer/chip or across multiple wafers/chips using a minimal number of samples~\cite{spatialvariation_bi_yuan2008spatial,spatialvariation_yuan2010bayesian,spatialvariation_bmf_zhang2014bayesian,spatialvariation_virtualprobe_zhang2010bayesian,spatialvariation_gp_nathan2012spatial,spatialvariation_earliest_meyer1989modeling,spatialvariation_huang2013process,spatialvariation_yuan2008spatial,spatialvariation_stine1997analysis,spatialvariation_kong2020semisupervised,spatialvariation_em_sherief2009analyzing}. This minimal sample requirement is a consequence of the fact that conducting specification tests on a wafer can be both time-consuming and economically costly. For example, as stated in~\cite{spatialvariation_bayesmultiwafer_zhang2010multiwafer}, the procedure of chip measurement, such as wafer probe testing, can cause mechanical stress damage to the wafer being tested, and may also be time-consuming due to the limited availability of input/output ports. It is worth noting that the concept of analyzing spatial variations at the wafer level has been explored for quite some time, dating back to at least 1989 as demonstrated in~\cite{spatialvariation_earliest_meyer1989modeling}. Our review primarily focuses on recent papers utilizing a Bayesian approach to address this issue.

\begin{figure}
    \centering
    \includegraphics[width=0.7\linewidth]{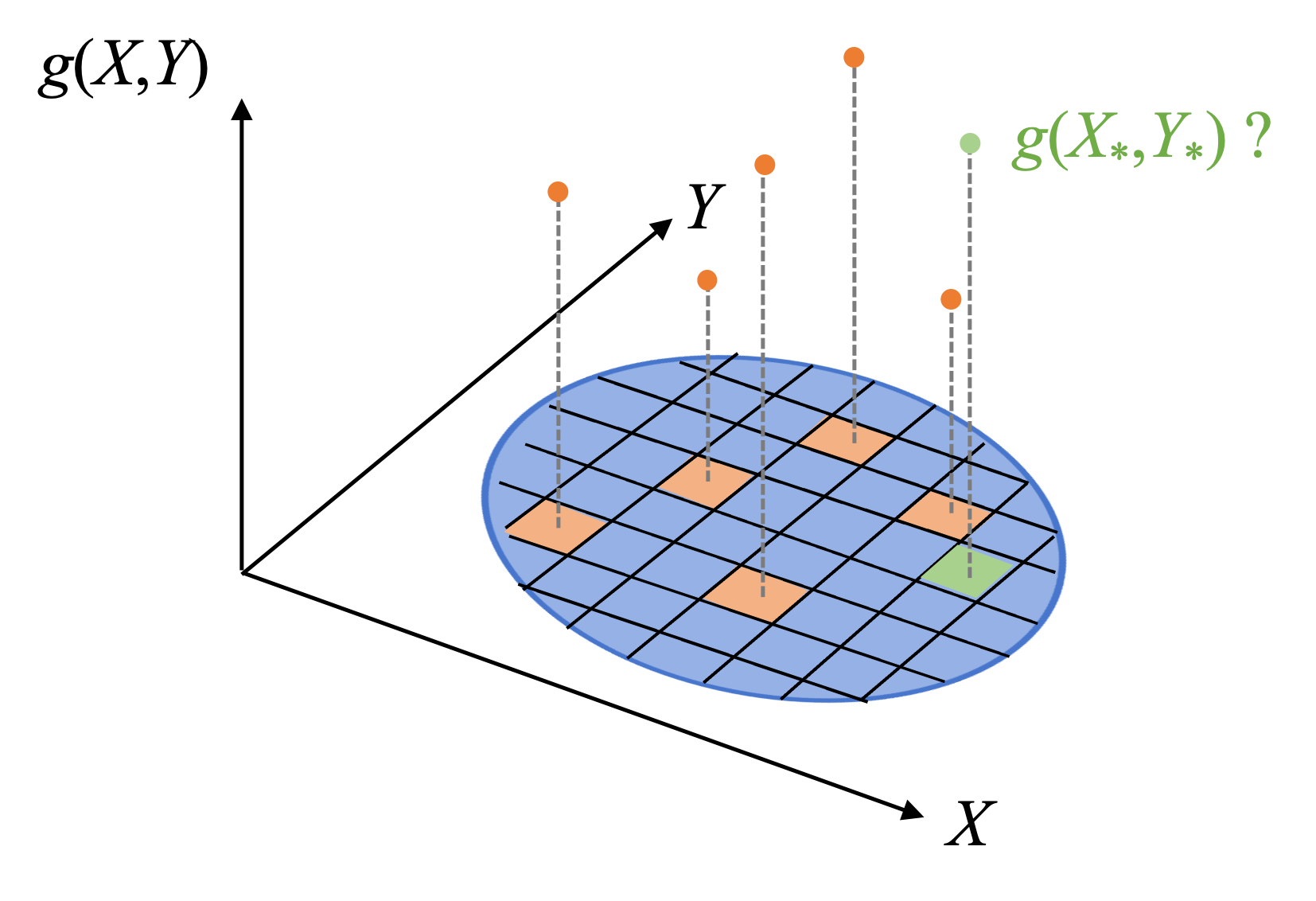}
    \caption{A visualization of the wafer-level spatial variation modeling problem. The performance values are known at the orange spots, and that of the green spot is required to be estimated.}
    \label{fig:spatialvaration_demo}
\end{figure}

Let us formally define the problem of spatial variation modeling. Notice that we usually deal with wafers in a two-dimensional Euclidean space. For ease of description, we use capitalized $X$ and $Y$ to represent the two axes. When considering a wafer-level problem, the previous Eq.~(\ref{eq:reparam}) can be rewritten more explicitly by considering the location dependence:
\begin{equation}
    \mathbf{z}(X,Y)=\mathbf{x}(X,Y)+\boldsymbol{\epsilon}(X,Y)\;.
\end{equation}
In essence, the above equation says that all the quantities $\{\mathbf{z},\mathbf{x},\boldsymbol{\epsilon}\}$ are now functions of the location $(X,Y)$. Consequently, the performance of interest $\mathbf{y}$ will also be a function of the location, because $\mathbf{y}=\mathbf{f}(\mathbf{z})$. For simplicity, we denote it by $\mathbf{y}=\mathbf{g}(X,Y)$.\footnote{Mathematically, $y=\mathbf{f}(\mathbf{z})=\mathbf{f}(\mathbf{z}(X,Y))=\mathbf{g}({X,Y})$, if we denote a composite function $\mathbf{g}=\mathbf{f}\circ\mathbf{z}$.} In wafer-level spatial variation modeling, we also care about one single performance in most cases; as before, we will use $\{g,y\}$ to replace $\{\mathbf{g},\mathbf{y}\}$ in such cases. Spatial variation modeling asks for the prediction of $\mathbf{y}$ at a given location $(X_\star,Y_\star)$, or equivalently to learn the function mapping $\mathbf{g}(\cdot,\cdot)$ based on a set of training data $\mathcal{D}=\{(X_i,Y_i,\mathbf{y}_i)\,|\mathbf{y}_i=\mathbf{g}(X_i,Y_i)\,,\,i=1,2,\cdots,N\}$. See Fig.~\ref{fig:spatialvaration_demo} for an illustration. The study of spatial variation problems can be divided into two main categories, based on whether the dataset $\mathcal{D}$ is provided or not. The first category assumes that $\mathcal{D}$ is given and the problem can be treated as a regression problem in a two-dimensional space. The second category~\cite{spatialvariation_virtualprobe_zhang2010bayesian} addresses the problem of how to acquire $\mathcal{D}$ (i.e., where to sample the training data) and learn the function mapping $\mathbf{g}(\cdot)$ simultaneously. In the following, we will detail a few papers that utilize Bayesian methods to address spatial variation modeling.

The Virtual Probe (VP) method~\cite{spatialvariation_vp_li2009virtual}, inspired by the principles of compressed sensing, utilizes Fourier transforms and maximum-a-posteriori (MAP) estimation to learn the mapping function $g(\cdot,\cdot)$. The authors employ a two-dimensional discrete Fourier transform (DCT) as a model template and aim to infer the Fourier coefficients from a given training dataset $\mathcal{D}$, so as to enable the recovery of performance at any arbitrary location through the application of inverse DCT~\cite{spatialvariation_vp_li2009virtual}. However, this leads to an under-determined least square problem that cannot be solved by the matrix inverse. To address this issue, the authors introduce a zero-mean Laplacian distribution for the Fourier coefficients, which is motivated by the measurement of real wafers. Subsequently, the MAP estimator can be calculated by solving an L1-norm regularization problem via linear programming. The authors demonstrate that the proposed method can achieve a significant reduction of approximately $10\times$ in error when compared to interpolation through the examination of two industrial examples. 

The authors further extend their work in~\cite{spatialvariation_virtualprobe_zhang2010bayesian,spatialvariation_bayesmultiwafer_zhang2010multiwafer}. In~\cite{spatialvariation_virtualprobe_zhang2010bayesian}, two major improvements are made to the original VP method. Firstly, the prior is defined as a multivariate Gaussian distribution, which allows for the fast evaluation of the posterior through the use of analytical forms. Secondly, an optimal sampling scheme, based on differential entropy, is proposed. This extension is referred to as the Bayesian Virtual Probe (BVP) and it demonstrates a $1.5\times$ reduction in error compared to the original VP method. In~\cite{spatialvariation_bayesmultiwafer_zhang2010multiwafer}, the authors expand the VP method to the multi-wafer context, and the Multi-wafer Virtual probe (MVP) method is proposed. MVP takes physical samples from different locations on each wafer and incorporates wafer-to-wafer correlation into the prior distribution. This enables the simultaneous estimation of spatial variations across all wafers through the use of the MAP estimator. Additionally, a robust regression algorithm is developed to accurately detect and remove measurement outliers. MVP demonstrates superior accuracy compared to the original VP method and the expectation-maximization method~\cite{spatialvariation_em_sherief2009analyzing}. 

The authors in~\cite{spatialvariation_bayestensor_luan2020prediction} use the tensor approach to further reduce the computational cost of VP. Their key is to estimate the data of multiple dies simultaneously by performing tensor completion in a higher-dimensional data space, which is realized by a variational Bayesian approach. 

In~\cite{spatialvariation_gp_nathan2012spatial}, the authors propose using a Gaussian process regression (GPR) model to learn the mapping function $g(\cdot,\cdot)$. The authors use several implementation techniques such as regularization by additive noise and leave-one-out cross-validation to effectively train the GPR model. They demonstrate that the GPR model outperforms the VP method~\cite{spatialvariation_vp_li2009virtual} on average by $0.5\times$ with much less run-time using industrial high-volume semiconductor manufacturing data. Later, in~\cite{spatialvariation_kmeans_ke2013handling}, the same authors acknowledge that most proposed methods, such as GPR and VP, assume that the target mapping $g(\cdot,\cdot)$ is continuous and smooth. However, actual production data often exhibits localized spatial discontinuous effects. To address this issue, the authors propose first partitioning a wafer into several clusters based on discontinuous effects using the K-means algorithm and then constructing individual GPR models within each cluster. This method demonstrates superior accuracy compared to the original GPR and VP methods on an RF transceiver example.

The authors in references~\cite{spatialvariation_bi_yuan2008spatial,spatialvariation_yuan2010bayesian,spatialvariation_hierarchibayes_yuan2011yield,spatialvariation_waferdefect_hwang2007model} have examined the problem of recognizing patterns of defects at the wafer level. In their work, the function $g(X,Y)$ is a probability density function that denotes the likelihood of observing a defect at location $(X,Y)$. We consider~\cite{spatialvariation_waferdefect_hwang2007model} as an example. If there are $K$ causes of the defect, then $g(X,Y)$ can be mathematically represented by a mixture model $g(X,Y)=\sum_{k=1}^K \pi_k p(X,Y|\boldsymbol{\theta}_k)$, where $0\leq\pi_k\leq 1$ is the $k$-th weight ratio and $\boldsymbol{\theta}_k$ is the parameter for the $k$-th defect generation principle~\cite{spatialvariation_waferdefect_hwang2007model}. For instance, when one manufacturing process causes the defect distributed uniformly on the wafer, we can use a component $p(X,Y|\boldsymbol{\theta}_1)=1/D$ to represent it, where $D$ represents the area of the wafer~\cite{spatialvariation_waferdefect_hwang2007model}. After the forms of $p(X,Y|\boldsymbol{\theta}_k)$ for all $k=1,2,\cdots,K$ are specified, the expectation-maximization algorithm can be used to calculate all $\{\pi_k,\boldsymbol{\theta}_k\}$ through maximizing the likelihood function~\cite{spatialvariation_waferdefect_hwang2007model}. Next, principle curves of defects can be drawn and defect catch and error rate can be calculated to evaluate the efficacy of the algorithm. Through simulated examples, it is demonstrated that the proposed method is effective in recovering the defect shape and attaining a small error rate and a large catch rate. For more implementation details, readers can refer to references~\cite{spatialvariation_bi_yuan2008spatial,spatialvariation_yuan2010bayesian,spatialvariation_hierarchibayes_yuan2011yield,spatialvariation_waferdefect_hwang2007model}. It should be noted that these works primarily differ in the specific forms of $p(X,Y|\boldsymbol{\theta}_k)$ and the solving algorithm. 

There are a few additional relevant works on spatial variation analysis using Bayesian approaches. The purpose of~\cite{spatialvariation_sdp_liu2018wafer} is to establish a statistical process control system for monitoring non-normal wafer thickness profiles in an industrial slicing process. The authors propose a mixed-effect profile monitoring approach based on the Dirichlet process, which demonstrates superior performance compared to benchmark methods and effectively detects deviant wafers with a minimum average rate of missed detection. In~\cite{spatialvariation_huang2013process}, the authors propose to use the K-means clustering algorithm to break down the systematic wafer-level variation into a set of weighted spatial basis functions. It achieves perfect matches with golden results and superior accuracy compared to a baseline method. In~\cite{spatialvariation_vae_hwang2020}, the authors argue that the clustering of wafer maps has become more difficult due to the complex patterns and high-dimensional data of wafer maps. They propose a Gaussian mixture model and a Dirichlet process within a variational autoencoder framework for efficient wafer clustering. The method is demonstrated on a real dataset from a global semiconductor manufacturing company and outperforms Bayesian nonparametric methods.

\subsection{Other Uses and Discussions}

In addition to the previously discussed circuit problems, Bayesian methods have also been utilized in several other modeling, diagnosis, and analysis problems. As these topics do not warrant a dedicated section individually, we provide a comprehensive summary for them in Table~\ref{tab:bayes_modeling_others}. Readers interested in additional information on these subjects can refer to the cited references in the table. It is important to note that the categorization in Table~\ref{tab:bayes_modeling_others} and this section are not mutually exclusive, and a paper may belong to multiple categories. For instance, the paper~\cite{processcontrol_yang2012bayesian} addresses fault diagnosis and prognosis in the chemical vapor deposition (CVD) process, thus it is suitable for both the ``fault diagnosis'' and the ``process control'' categories. Similarly, the paper~\cite{fault_litho_villacourt1993designing} encompasses both ``fault diagnosis'' and ``lithography analysis.'' We reiterate that the categorization of each paper is based on the primary keywords specified in the original paper.

\begin{table*}[!htb]
    \centering
    \caption{A summary of papers on Bayesian circuit modeling, analysis, and diagnosis beyond those discuss in the main text}
    \begin{threeparttable}
    \begin{tabular}{c|c|c|c|c}
    \toprule
    Task & Paper & Method Keyword & Major Test Object & Results Summary \\
    \midrule
     \multirow{12}{*}{\thead{Fault \\ diagnosis}} 
       & \cite{fault_mittelstadt1995application} & Belief network & Contact resistance & Results of $3$ logged trails matched with the expected \\
       & \cite{fault_brandt1997circuit} & BNN & Differential amplifier & $<1\%$ prediction error over $60\%$ test cases \\
       & \cite{fault_aminian2001fault}   & BNN + preprocessing & Analog filter & $96\%$ correct classification of test data \\
       & \cite{fault_liu2006parametric} &   Bayes' theorem & Two-stage Op-Amp &  $86\%$ of process/layout faults diagnosed\\
       & \cite{fault_bn_Jha2009localizing} & Belief network & ISCAS'89 benchmark circuits & Transient error diagnosed \\
       & \cite{fault_alsoprocesscontrol_li2013fault} & Belief network & Chipset assembly, test factory &  Good agreement with real data \\
       & \cite{fault_nawaz2014fault} & Belief network & Etch equipment & 4 out of 5 faults identified correctly\\
       & \cite{fault_wang2017novel} & Belief network  & A wafer dataset & Over $90\%$ accuracy in most cases \\
       & \cite{fault_khakifirooz2018bayesian}  & Gibbs sampling & 20 lots of 500 wafers & Smaller error compared to baselines \\
       & \cite{fault_bayes_he2019feature}  & Variational Bayes & Chaotic circuit, Op-Amp filter & $>1\%\uparrow$ accuracy compared to baselines\\
       & \cite{fault_nb_he2019naive}  & Naive Bayes classifier & Chaotic circuit, Op-Amp filter & $0.2\%,2.5\%,3.2\%\uparrow$ accuracy compared to baselines\\
       & \cite{fault_fu2022bayesian} & Belief network & Probe card & About $3\times$ accuracy improvement and $64\%$ cost reduction \\

      \midrule
      \multirow{10}{*}{\thead{Reliability \\ analysis}}
      & \cite{fault_litho_villacourt1993designing} & MAP & A lithography tool & Test length shortened \\ 
      & \cite{reliability_krishnaswamy2005accurate} & PTM & logic circuits & $>5\%\uparrow$ reliability improved \\
      & \cite{reliability_choudhury2009reliability} & Bayes' theorem $^1$ & logic circuits & $<12\%\downarrow$ error compared to MC w/ less run-time\\
      & \cite{reliability_ibrahim2011using} & Belief network & CMOS gates & Better accuracy compared with one baseline \\
      & \cite{reliability_khalid2011reliability} & Belief network & 5 logic circuits & $>2\%\uparrow$ accuracy compared to baselines \\
      & \cite{reliability_muratovic2013modelling} & MAP & HV circuit breaker & Reliability value obtained\\
      & \cite{reliability_ibrahim2014accurate} & Belief network $^2$  & ISCAS'85 benchmark circuits &  Faster and more accurate compared to exact inference\\
      & \cite{reliability_jiang2016reliability} & Belief network & ZPW-2000A track circuits & Reliability value obtained \\
      & \cite{reliability_vallero2016cross} & Belief network & Microprocessors & More accurate FIT compared to FI w/ $65\%\downarrow$ run-time \\
      & \cite{reliability_rabiei2018component} & Belief network & COTS devices & Failure rate calculated \\
      
      \midrule
      \multirow{8}{*}{\thead{Process\\ control}}
      & \cite{processcontrol_yousry1991process} & Empirical Bayes & Printed circuit board & Deployed in AT\&T factories\\
      & \cite{processcontrol_rao1996monitoring}  & MAP & VLSIC process & Detection time reduced for a multi-stage process\\
      & \cite{processcontrol_kong2010process} & Particle filter & CMP process (MRR prediction) & $13\%\uparrow$ improvement compared to baselines \\
      & \cite{processcontrol_kong2011nonlinear} & Particle filter & CMP process (end-point detection) & Good agreement with real data \\
      & \cite{processcontrol_yang2012bayesian} & Belief network & CVD process &  Around $90\%$ classification accuracy \\
      & \cite{processcontrol_yu2012semiconductor} & Gaussian mixture & Aluminum stack etch process &  Outperforms PCA-based monitoring models\\
      & \cite{processcontrol_chen2019physics} & BNN + physical prior & Aspect ratio dependent etch & Better accuracy compared to a baseline BNN \\
      
      \midrule
      \multirow{5}{*}{\thead{Parameter \\ extraction}}
      & \cite{extraction_ci_sharma1993optima} & Confidence region & MOSFET & Good agreement with real data \\
      & \cite{extraction_rtn_awano2013multi}  & Gibbs sampling & multi-trap RTN & $2.1\times\uparrow$ accuracy compared to one baseline (HMM) \\
      & \cite{extraction_bayes_Li2016compact}  & MAP & MOSFET & Outperforms baselines (LSE and BPV)\\
      & \cite{extraction_library_Li2015statistical} & MAP & Delay and slew of standard cell & $15\times\downarrow$ run-time compared with baselines \\
      & \cite{extraction_hemt_cai2018bayesian} & MAP & GaN HEMT & Good agreement with real data\\

      \midrule
      \multirow{4}{*}{\thead{Timing \\ analysis}}
      & \cite{timing_bayes_bhardwaj2003tau} & Belief network & ISCAS'85 benchmark circuits &  $\sim 70\%$ circuit size reduction w/o loss of accuracy \\
      & \cite{timing_bi_lee2006refined} & MAP & Several benchmark circuits & Spatial delay correlations learned \\
      & \cite{timing_svm_wang2007design} & SVM$^3$ & Microprocessors & Correlation in path delay learned\\
      & \cite{timing_review_blaauw2008statistical} & A survey paper & Digital circuits & Various methods reviewed (e.g., Bayesian methods) \\ 

      \midrule
      \multirow{4}{*}{\thead{Lithography \\ analysis}}
      & \cite{hotspot_matsunawa2016laplacian} & MAP & 32nm industrial layouts & Outperforms a few baselines \\
      & \cite{hotspot_zhang2016enabling} & Online naive Bayes &  ICCAD'12 benchmark layouts & $3.47\%\uparrow$ accuracy compared to baselines w/ $58.88\%\downarrow$ cost \\
      & \cite{hotspot_alawieh2021adapt} & MAP & ICCAD'12 benchmark layouts & Outperforms several baselines \\
      & \cite{hotspot_ye2019litho} & Gaussian process & ICCAD'12 benchmark layouts & Comparable accuracy to baselines w/ $28\%\downarrow$ in false alarms \\
    \bottomrule
    \end{tabular}
    \label{tab:bayes_modeling_others}
    \begin{tablenotes}
    \footnotesize
    \item $^1$: Precisely, it uses the product rule of probability.
    \item $^2$: Belief network presented in the third step of the proposed CSA approach. It shows that overall CSA outperforms exact inference on a Belief network.
    \item $^3$: SVM is short for support vector machine. It is shown in~\cite{bayestheory_svm_sollich2002bayesian} that SVM is the MAP estimator to inference problems with Gaussian process priors.
    \item $^\dagger$ Note: MC $=$ Monte Carlo. Refer to the corresponding papers for other unexplained abbreviations.
    \end{tablenotes}
    \end{threeparttable}
\end{table*}

In reviewing these papers, we note the similarities between yield and reliability problems, while also acknowledging their differences. Yield refers to the percentage of functioning devices at the point of manufacturing, while reliability pertains to the percentage of functioning devices during or over subsequent use conditions after the chip is fabricated. However, some current literature does not fully distinguish between these two concepts and tends to blend their use. To ensure clear communication in the field, we suggest that authors use these terms with precision and accuracy, avoiding blurring their distinct meanings. Careful and accurate use of nomenclature will help prevent confusion and facilitate progress in this important area of research.

Second, deep neural networks have achieved remarkable success in recent years, and we see a growing interest in the literature to use them to address circuit modeling, diagnosis, and analysis problems as well. Although our focus in this review is on Bayesian methods, it is important to discuss the advantages and disadvantages of both approaches. Deep neural networks have the ability to learn intricate and non-linear relationships between inputs and outputs, making them well-suited for tasks with high-dimensional and non-linear data. However, they are often computationally expensive and require a large amount of data to train effectively. For instance, ImageNet, which contains over 14 million images, is commonly used to train deep neural networks for computer vision tasks. However, obtaining such large datasets can be challenging in the field of EDA due to IP constraints and simulation/measurement costs, among other factors. Alternatively, Bayesian methods may provide more reliable results with smaller datasets and are less computationally expensive. They are also simpler, and easier to understand and interpret, but may not be suitable for complex problems. Ultimately, the choice of which method to use depends on the specific problem and available resources, and we alert the readers that deploying deep neural networks might not be a panacea.
\section{Optimization}\label{sec:optimization}

Historically, adopting optimization algorithms to assist circuit design thrived in the latter half of the 20th century as computers became more powerful~\cite{reviewoptimization_brayton1981survey}. While the basic problem formulation remains unchanged, a variety of new and effective optimization techniques have emerged. Among these methods, Bayesian optimization (BO) stands out as particularly attractive in recent decades~\cite{bocircuitopt_touloupas2021locomobo,bocircuitopt_touloupas2021local,bocircuitopt_liu2021parasitic,bocircuitopt_lyu2017efficient,bocircuitopt_lyu2018multi,bocircuitopt_zhang2021efficient,bocircuitopt_touloupas2022mixed,bocircuitopt_zhang2020efficient,bocircuitopt_liao2021high,bocircuitopt_zhang2019bayesian,bocircuitopt_chen2022high,bocircuitopt_lyu2018batch,bocircuitopt_he2022batched,bocircuitopt_he2020efficient,bocircuitopt_abdelaal2020bayesian,bocircuitopt_huang2021bayesian,bocircuitopt_vicsan2022automated,bocircuitopt_touloupas2022mixed2,bocircuitopt_fu2022batch,bocircuitopt_wang2022analog,bocircuitopt_lu2021automated,bocircuitopt_huang2021robust,bocircuitopt_zhang2019efficient,bocircuitopt_zhang2022lineasybo,bocircuitopt_huang2022bayesian,bocircuitopt_lu2020mixed,bocircuitopt_wang2021high,bocircuitopt_zhao2022novel,bocircuitopt_yin2022asynchronous,bocircuitopt_yin2022fast,bocircuitopt_gao2019efficient,bocircuitopt_yin2022efficient,bocircuitopt_pan2019analog,bocircuitopt_touloupas2021optimization}. This section will review the utilization of BO for a diverse range of optimization challenges in circuit design.


\subsection{Schematic-level Circuit Optimization}

The circuit optimization problem considered throughout this section is at the schematic level. This problem is also known as circuit synthesis or circuit inverse design in the literature. Considering the type of circuit being optimized, this problem can be broadly categorized into two classes, analog circuit optimization, and digital circuit optimization. We start this subsection with analog circuit optimization.

Adhering to the notation introduced in Section~\ref{sec:preliminary}, the basic form of analog circuit optimization disregards the stochastic variation $\boldsymbol{\epsilon}$ (i.e., assumes $\boldsymbol{\epsilon}=\mathbf{0}$) and aims to minimize a scalar circuit performance metric through the optimization of design variables:
\begin{equation}~\label{eq:basic_formulation_opt}
    \min_{\mathbf{x}\in\Gamma} f(\mathbf{z})=f(\mathbf{x})\quad\text{($\boldsymbol{\epsilon}=\mathbf{0}$)}\;,
\end{equation}
where $\mathbf{x}$ represents the design variables, $\Gamma$ represents the feasible design space, and $f(\cdot)$ represents the mapping from design variables to the scalar circuit performance of interest (e.g., gain or phase margin of an Op-Amp). The most commonly considered analog circuit optimization problem is analog sizing. In its strict definition, this refers to the scenario in which $\mathbf{x}$ represents the width and/or length of the transistors in an analog circuit, so enlarging or shrinking $\mathbf{x}$ corresponds to sizing the transistor. Typically, $\mathbf{x}$ is a continuous variable (e.g., $\mathbf{x}\in\Gamma\subseteq\mathbb{R}^H$) in analog sizing, and as such, the classical BO shown in Algo.~\ref{algo:bayesopt} and its variants can be applied. Alternatively, in a general analog circuit optimization problem, $\mathbf{x}$ might be discrete or even mixed discrete-continuous. As an example, \cite{bocircuitopt_touloupas2022mixed} includes the number of turns in a 90~nm spiral inductor and the number of fingers in a 90~nm metal capacitor as design variables, both of which are discrete. When different aspects of an analog circuit optimization problem are emphasized (e.g., process variation considered), the formulation shown in Eq.~(\ref{eq:basic_formulation_opt}) might need to be modified accordingly, and the BO algorithm need to be refined. In the following, we will outline these extensions, and then classify the relevant literature based on the extension(s) they address. 

The first extension is taking into consideration the stochastic variation represented by $\boldsymbol{\epsilon}\sim p(\boldsymbol{\epsilon})$, referred to as variation-aware or robust analog circuit optimization. This problem is formulated in various ways in the literature due to implementation differences, and here we list three commonly used ones. The first formulation directly optimizes the expected value of the performance $f(\mathbf{z})$: 
\begin{equation}\label{eq:expected_opt}
\begin{aligned}
\min_{\mathbf{x}\in\Gamma} \mathbb{E}[f(\mathbf{z})]&=\mathbb{E}[f(\mathbf{x}+\boldsymbol{\epsilon})]\quad \text{where} \ \boldsymbol{\epsilon}\sim p(\boldsymbol{\epsilon})\\
&\approx \frac{1}{N}\sum_{n=1}^N f(\mathbf{x}+\boldsymbol{\epsilon}_n)\;.\\
\end{aligned}
\end{equation}
Note that the expectation is performed with respect to $\boldsymbol{\epsilon}$, and it can be approximated by $\sum_{n=1}^N f(\mathbf{x}+\boldsymbol{\epsilon}_n)/N$, where $\boldsymbol{\epsilon}_n$ is a sample drawn from $p(\boldsymbol{\epsilon})$. It is straightforward as an extension of Eq.~(\ref{eq:basic_formulation_opt}) to the case when randomness is present. 

Alternatively, a second formulation requires an acceptable performance region $\Omega$, and minimizes the probability that the performance $f(\mathbf{z})$ falls outside of $\Omega$:
\begin{equation}\label{eq:yield_opt}
\begin{aligned}
\min_{\mathbf{x}\in\Gamma} P[f(\mathbf{z})\notin\Omega]&=P[f(\mathbf{x}+\boldsymbol{\epsilon})\notin\Omega]\quad (\boldsymbol{\epsilon}\sim p(\boldsymbol{\epsilon}))\\
&\approx\frac{1}{N}\sum_{n=1}^N [[f(\mathbf{x}+\boldsymbol{\epsilon}_n)\notin\Omega]] \;,\\
\end{aligned}
\end{equation}
where $[[A]]$ represents the Iverson braket, equal to one when the expression $A$ holds true, and zero otherwise. Refer to Eq.~(\ref{eq:rarefailure_expression2}) for more background details. The above formulation is called yield optimization, as it attempts to minimize the failure rate, or equivalently, maximize the yield. 

The last formulation attempts to improve the worst-case performance by solving:
\begin{equation}\label{eq:worst_case_opt}
\begin{aligned}
\min_{\mathbf{x}\in\Gamma}\max_{\boldsymbol{\epsilon}} f(\mathbf{z})&=\min_{\mathbf{x}\in\Gamma}\max_{\boldsymbol{\epsilon}}f(\mathbf{x}+\boldsymbol{\epsilon})\quad (\boldsymbol{\epsilon}\sim p(\boldsymbol{\epsilon}))\\
&\approx \min_{\mathbf{x}\in\Gamma}\max \{f(\mathbf{x}+\boldsymbol{\epsilon}_n)\,|\,n=1,\cdots,N\}.\\
\end{aligned}
\end{equation}
Note that $\max_{\boldsymbol{\epsilon}}f(\mathbf{x}+\boldsymbol{\epsilon})$ represents the worst-case performance at $\mathbf{x}$ when variation is considered, and it can be approximated by $\max \{f(\mathbf{x}+\boldsymbol{\epsilon}_n)\,|\,n=1,2,\cdots,N\}$, where $\boldsymbol{\epsilon}_n$ is a sample drawn from $p(\boldsymbol{\epsilon})$. 

These three formulations have their own strengths and weaknesses. Yield optimization and worst-case optimization both may result in overly conservative designs. On the other hand, expected value optimization may not reflect the worst-case scenario. The choice of which formulation to use depends on the specific problem requirements and user preference. 

Finally, we emphasize that advanced BO approaches exist for solving Eq.~(\ref{eq:expected_opt})-(\ref{eq:worst_case_opt}) without explicitly using the approximations shown in the second lines of those equations, such as StableOpt~\cite{bo_stableopt_bogunovic2018adversarially} for Eq.~(\ref{eq:worst_case_opt}). Alternatively, when those approximations are used, Eqs.~(\ref{eq:expected_opt})-(\ref{eq:worst_case_opt}) are reduced into the form of Eq.~(\ref{eq:basic_formulation_opt}) and the classical BO can be applied. To understand this, we can view the expression $\sum_{n=1}^N f(\mathbf{x}+\boldsymbol{\epsilon}_n)/N$ in the case of Eq.~(\ref{eq:expected_opt}) as a new function $f(\mathbf{x})$ with a slight abuse of notation. This treatment is frequently used in analog circuit optimization over multiple process-voltage-temperature (PVT) corners~\cite{bocircuitopt_touloupas2021locomobo}. Namely, in these works, $\boldsymbol{\epsilon}_n$ represents the equivalent variation introduced by the $n$-th PVT corner and evaluating $f(\mathbf{x}+\boldsymbol{\epsilon}_n)$ will invoke the circuit simulator at the $n$-th PVT corner. These works usually invoke the circuit simulator $N$ times, each time at one PVT corner, and take the average performance (Eq.~(\ref{eq:expected_opt})),  yield (Eq.~(\ref{eq:yield_opt})), or worst-case performance (Eq.~(\ref{eq:worst_case_opt})) as the ultimate objective desired to optimize. In our review, these papers will be considered variation-aware optimization, though some of them might not explicitly claim this.

In a second extension, constraints are introduced into the formulation to ensure the some auxiliary circuit performances are met when optimizing one user-defined predominant performance. Taking Eq.~(\ref{eq:basic_formulation_opt}) as an example, besides the goal of minimizing $f(\mathbf{x})$, suppose we have $M$ additional circuit performances which we want to constrain to be within certain bounds. Then the following constrained optimization can be used:
\begin{equation}~\label{eq:constrained_formulation_opt}
 \begin{aligned}
    &\min_{\mathbf{x}\in\Gamma} f(\mathbf{z})=f(\mathbf{x})\quad \text{where}\ \boldsymbol{\epsilon}=\mathbf{0}\\
    &\text{s.t.,} \quad L_i \leq g_i(\mathbf{x}) \leq U_i\quad i=1,2,\cdots,M \;;\\
 \end{aligned}   
\end{equation}
where $g_i(\mathbf{x})$ represents the $i$-th auxiliary circuit performance, and $\{L_i,U_i\}$ represents its acceptable lower and upper bound, respectively. For instance, in the power amplifier example studied by~\cite{bocircuitopt_lyu2017efficient}, $f(\mathbf{x})$ symbolizes the output efficiency, while $g_1(\mathbf{x})$ and $g_2(\mathbf{x})$ denote the output power and total harmonic distortion, respectively. 

Constraints can be similarly added to Eq.~(\ref{eq:expected_opt})-(\ref{eq:worst_case_opt}). BO has already been extended to deal with such a constrained optimization problem~\cite{bo_constraint_gardner2014bayesian,bo_constraint_eriksson2021scalable,bo_constraint_letham2019constrained}, such as by using a good acquisition function~\cite{bo_constraint_gardner2014bayesian}. We emphasize that $\mathbf{x}\in\Gamma$ and $\{L_i \leq g_i(\mathbf{x}) \leq U_i\,|\,i=1,2,\cdots,M\}$ are usually used to denote two different types of constraints, motivating our use of separate notations. Specifically, $\mathbf{x}\in\Gamma$ represents the permissible range of the design variable (e.g., the width of a transistor is in $[10\text{nm},20\text{nm}]$), and it has almost no cost to evaluate whether $\mathbf{x}$ is in $\Gamma$ or not. In contrast, $\{L_i \leq g_i(\mathbf{x}) \leq U_i\,|\,i=1,2,\cdots,M\}$ might further shrink the region of $\Gamma$, and circuit simulations are needed to evaluate whether a design variable $\mathbf{x}$ satisfies these constraints on circuit performances. In our review here, constrained circuit optimization always refers to the fact that constraints are imposed on circuit performances, i.e., $\{L_i \leq g_i(\mathbf{x}) \leq U_i\,|\,i=1,2,\cdots,M\}$.

\begin{figure}[!htb]
    \centering
    \includegraphics[width=0.35\textwidth]{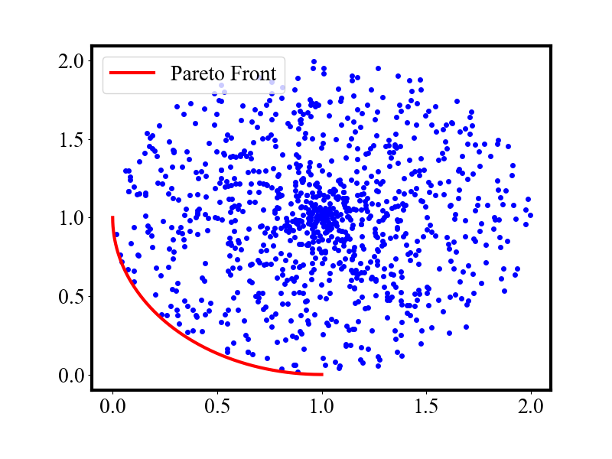}
    \caption{Illustration of a Pareto front. The X-axis and Y-axis present the value of $f_1(\mathbf{x})$ and $f_2(\mathbf{x})$, respectively. We randomly sample a large number of designs from the feasible domain $\Gamma=\{\mathbf{x}\,|\,0\leq x_1,x_2,x_3\leq 1\}$ and scatter their associated performance value $\mathbf{f}(\mathbf{x})$ to the two-dimensional space.}
    \label{fig:mo_demo}
\end{figure}

The third extension applies to the case when multiple performances are of interest, where another formulation is used rather than treating one performance as predominant and the others as auxiliary. Namely, we treat all performances equally as objectives and directly optimize a vector function $\mathbf{f}(\mathbf{x})$ as opposed to a scalar function $f(\mathbf{x})$. Given that conflicting objectives are often present, a single optimum is usually unattainable, and instead, a set of solutions representing a trade-off between the conflicting objectives is sought. These solutions are referred to as Pareto optimal designs, and their performance values form a Pareto front. As an illustration, consider a three-dimensional design variable $\mathbf{x}=[x_1,\,x_2,\,x_3]^T$ where each $x_i\in[0,1]$, and $\mathbf{f}(\mathbf{x})=[f_1(\mathbf{x}),f_2(\mathbf{x})]^T$, where $f_1(\mathbf{x})=x_3\cos(\pi(x_1+x_2))+1$ and $f_2(\mathbf{x})=x_3\sin(\pi(x_1+x_2))+1$. The aim is to minimize $\mathbf{f}(\mathbf{x})$ with respect to $\mathbf{x}$. The minimum value of $f_1(\mathbf{x})$ is $0$ when neglecting $f_2(\mathbf{x})$, and similarly, the minimum value of $f_2(\mathbf{x})$ is $0$ when neglecting $f_1(\mathbf{x})$. However, it is impossible for both $f_1(\mathbf{x})$ and $f_2(\mathbf{x})$ to reach their minimum values simultaneously with a single $\mathbf{x}$, as illustrated in Fig.~(\ref{fig:mo_demo}). Blue solid dots located at the bottom left in Fig.~(\ref{fig:mo_demo}) represent better designs. However, the form of $\mathbf{f}(\mathbf{x})$ limits the ability to move infinitely towards the bottom left, and the red curve displays the limit of this movement, or equivalently, how well we can minimize both $f_1(\mathbf{x})$ and $f_2(\mathbf{x})$. Designs whose performance values fall on the red curve are considered Pareto optimal, as they cannot be improved in one objective without negatively impacting the other objectives. This problem is known in the literature as multi-objective optimization, many-objective optimization, or Pareto front modeling, with subtle differences among the terms.\footnote{Multi-objective optimization refers to two or more objectives, while many-objective optimization is a type of multi-objective optimization but with additional challenges such as a large number of objectives. Both of these terms care about a Pareto optimal solution $\mathbf{x}$ in the design space. On the other hand, Pareto front modeling emphasizes the parameterization of the Pareto front in the performance space and does not necessarily require a Pareto optimal solution $\mathbf{x}$.} 

Multi-objective BO has been extensively investigated in the current literature for such problems. Unlike conventional single-objective BO, multi-objective BO requires a more sophisticated surrogate model and acquisition function. For example, the commonly used surrogate models now include Multi-task GPR or Multi-output GPR. Additionally, various techniques have been proposed for the design of the acquisition function, including hypervolume improvement and nondominated sorting genetic algorithms, among others. For further information, interested readers may refer to~\cite{bo_multibo_hernandez2016predictive,bo_multibo_khan2002multi,bo_multibo_laumanns2002bayesian} and their cited references.

It is important to note that these three formulation enhancements (i.e., variation-aware/robust, constrained, and multi-objective) are not mutually exclusive. Indeed, a circuit optimization problem can adopt a formulation incorporating all of these enhancements simultaneously. With that in mind, we categorize the published papers on BO for analog circuit optimization based on which enhancements are considered in Table~\ref{tab:bayes_opt_paper}. Two points are noteworthy here. Firstly, as the major numerical results of almost all papers shown in Table~\ref{tab:bayes_opt_paper} can be concluded as outperforming some baselines, the description of results has been omitted. Secondly, various papers have applied specialized techniques to tackle specific challenges (e.g. high-dimensional BO, mixed-variable BO) or to enhance efficiency (e.g. asynchronous BO, multi-fidelity BO). In recognition of these specialties, a note has been included in the last column of Table~\ref{tab:bayes_opt_paper}.

\begin{table*}[!htb]
    \centering
    \caption{A summary of papers on Bayesian optimization for analog circuit optimization}
    \begin{threeparttable}
    \begin{tabular}{c|c|c|c|c|c|c}
    \toprule

    \multirow{2}{*}{Paper} & \multicolumn{3}{c|}{Formulation Enhancement} & Method & Test & Specialty \\

     & Robust & Constr & Multi-obj & Keyword &  Object & Summary \\

    \midrule
    \cite{bocircuitopt_liu2014gaussian} & \xmark & \cmark & \xmark & GPR + LCB & Power Amp (\conti-17), etc., & First few apply BO for analog sizing  \\
    \cite{bocircuitopt_lyu2017efficient} & \cmark & \cmark & \cmark & GPR + WEI & Amp-1,2 (\conti-5,\conti-24), etc., & First few apply BO for analog sizing \\
    \cite{bocircuitopt_lyu2018multi} & \cmark & \xmark & \cmark & GPR + LCB & Amp-1,2 (\conti-5,\conti-24), etc., & NSGAII solves multi-objective LCB \\
    \cite{bocircuitopt_lyu2018batch} & \xmark & \xmark & \xmark & GPR + Acq ensemble & Amp-1,2 (\conti-10,\conti-12) & Batch BO enabled by the ensemble \\
    \cite{bocircuitopt_zhang2019bayesian} & \cmark & \cmark & \xmark & GPR + WEI & Op-Amp (\conti-10), CP (\conti-36) &  Neural network as GPR kernel \\
    \cite{bocircuitopt_gao2019efficient} & \cmark & \xmark & \cmark & BNN + LCB & CP (\conti-16), Amp (\conti-13) & BNN as surrogate model \\
    \cite{bocircuitopt_zhang2019efficient} & \xmark & \cmark & \xmark & GPR + WEI & Amp (\conti-5), CP (\conti-36) & Multi-fidelity BO \\
    \cite{bocircuitopt_pan2019analog} & \cmark & \cmark & \xmark & GPC + its Eq.~(45) & Op-Amp(\conti-4), LNA (\conti-4) & Handles binary testing outputs \\
    \cite{bocircuitopt_lu2020mixed} & \xmark & \cmark & \xmark & GPR + WEI & CP (\conti-18, \discr-5), etc., & Modified kernel for \discr~variables  \\
    \cite{bocircuitopt_zhang2020efficient} & \xmark & \cmark & \cmark & GPR + UCB & Ampifier-1,2 (\conti-10,\conti-12) & Asynchronous BO \\
    \cite{bocircuitopt_he2020efficient} & \cmark & \cmark & \xmark & SP-GPR + WEI & Amp (\conti-24), VCO (\conti-20) & Sparse GPR model \\
    \cite{bocircuitopt_abdelaal2020bayesian} & \xmark & \cmark & \xmark & GPR + EI & Op-Amp (variable unknown) &  Classical BO  \\
    \cite{bocircuitopt_huang2021bayesian} & \xmark & \cmark & \xmark & MT-GPR + WEI  & Op-Amp (\conti-10), LNA (\conti-10) & MT-GPR learns multi-performances \\
    \cite{bocircuitopt_touloupas2021locomobo} & \cmark & \cmark & \cmark & GPR + TS & Amp-1,2,3 (\conti-23, \conti-43, \conti-21) & Local BO, trust-region method \\
    \cite{bocircuitopt_touloupas2021local} & \xmark & \cmark & \xmark & GPR + TS & Amp (\conti-23), LNA (\conti-33) & Local BO, sparse GPR model \\
    \cite{bocircuitopt_liu2021parasitic} & \cmark & \cmark & \xmark & GNN + WEI & Amp (\conti-2, \discr-8), driver (\conti-1, \discr-6) & Parasitic-aware, GNN w/ dropout \\
    \cite{bocircuitopt_zhang2021efficient} & \xmark & \cmark & \xmark & GPR + Acq ensemble & Amp-1,2,3 (\conti-10, \conti-12, \conti-24), etc., & Batch BO enabled by the ensemble\\
    \cite{bocircuitopt_liao2021high} & \cmark & \xmark & \cmark & Add-GPR + UCB & Amp, DC (\conti-na) & LDE-aware, high-dimensional  \\
    \cite{bocircuitopt_lu2021automated} & \xmark & \cmark & \xmark & GPR + WEI & Op-Amp (\conti-na) &  Bi-level BO, compensation design \\
    \cite{bocircuitopt_huang2021robust} & \xmark & \xmark & \xmark & GPR + EI & Amp-1,2 (\conti-10, \conti-12) & Local penalization~$^1$ \\
    \cite{bocircuitopt_touloupas2021optimization} & \xmark & \cmark & \xmark & GPR + modified TS &  Amp-1,2 (\conti-11, \conti-43) & Applied to technology migration \\
    \cite{bocircuitopt_yin2022efficient} & \cmark & \cmark & \cmark & Online GPR & Op-Amp-1,2 (\conti-11, \conti-21) & Self-adaptive incremental learning  \\
    \cite{bocircuitopt_zhao2022novel} & \cmark & \cmark & \xmark & GPR + wPESC & Amp-1,2 (\conti-10, \conti-11) & Automatically choose test benches \\
    \cite{bocircuitopt_yin2022asynchronous} & \cmark & \cmark & \cmark & GPR + EIM & Op-Amp-1,2 (\conti-11, \conti-21) & Asynchronous BO \\
    \cite{bocircuitopt_yin2022fast} & \cmark & \cmark & \cmark & Online GPR + EIM & Op-Amp-1,2 (\conti-11, \conti-26), etc., & Self-adaptive incremental learning \\
    \cite{bocircuitopt_zhang2022lineasybo} & \cmark & \cmark & \xmark & GPR + LCB/EI & CP (\conti-36), Amp (\conti-12) & Search in one-dimensional subspace \\
    \cite{bocircuitopt_huang2022bayesian} & \xmark & \cmark & \xmark & MT-GPR + EI & Transformer (\conti-4), LNA (\conti-15), etc., & Multitask NN as GPR kernel\\
    \cite{bocircuitopt_li2022performance} & \xmark & \cmark & \xmark & GPR + WEI & 5 OTAs, 2 VCOs, 2 SCFs (\discr-na), etc., & Wire sizing, GPR guided by GNN\\
    \cite{bocircuitopt_vicsan2022automated} & \cmark & \xmark & \cmark & GPR + LCB & Voltage regulator (\conti-17 + \discr-10), etc., & Novel evolutionary algorithm \\
    \cite{bocircuitopt_touloupas2022mixed2} & \cmark & \cmark & \xmark & GPR + TS & LNA (\conti/\discr-17) $^2$ & VAE converts \conti~variables to \discr \\
    \cite{bocircuitopt_wang2022analog} & \cmark & \xmark$^3$ & \xmark & GPR + ES & Comparator (\conti-12), LNA (\conti-13), etc., & Freeze-thaw BO \\
    \cite{bocircuitopt_chen2022high} & \cmark & \cmark & \xmark & GPR + EI & LDR (\conti-25), Amp (\conti-25), etc., & $g_m/I_D$ as variable, variable selection \\
    \cite{bocircuitopt_he2022batched} & \cmark & \cmark & \xmark & MF-GPR + WEI & CP (\conti-36), VCO (\conti-20), etc., & Multi-fidelity BO \\
    \cite{bocircuitopt_fu2022batch} & \xmark & \xmark & \xmark &  GPR+WEI & Op-Amp (\conti-24), CP (\conti-36) & Batch BO, multi-point selection \\
    \cite{bocircuitopt_touloupas2022mixed} & \xmark & \cmark & \xmark & GPR + TS & LNA (\conti-11 + \discr-10), etc., & VAE converts \conti~variables to \discr \\

    \bottomrule
    \end{tabular}
    \label{tab:bayes_opt_paper}
    \begin{tablenotes}
    \footnotesize
    \item $^1$ Although this paper has ``robust'' in the title, its robustness means its improved BO method can avoid bad designs at which circuit simulations fail.
    \item $^2$ This paper considers mixed-variable. However, it does not mention how many among the $17$ variables are continuous.
    \item $^3$ Its constraints (Eq (35)-(38)) actually define the region $\Omega$ given in our Eq.~(\ref{eq:yield_opt}), so it is an unconstrained analog circuit yield optimization.
    \item $^\star$ The presence of a checkmark in the columns marked ``Robust,'' ``Constr,'' or ``Multi-obj'' indicates that the corresponding work considers process variation (multi-corner scenario or layout-dependent effect), has constraints on circuit performances, or addresses multiple circuit performances, respectively.
    \item  $^\dagger$ In the ``Test Object'' column, the type and dimension of the design variable are indicated within the parentheses. The abbreviation ``\conti'' stands for ``continuous'' and ``\discr'' stands for ``discrete,'' with ``na'' indicating ``not available'.
    \item $^\ddagger$ In the ``Method keyword'' column, the type of surrogate model (such as GPR or BNN) and acquisition function (such as WEI or LCB) used in BO are recorded.
    \end{tablenotes}
    \end{threeparttable}
\end{table*}

To conclude this subsection, we consider the topic of digital circuit optimization. It is worth mentioning that the optimization of digital circuits has been a subject of research for many years~\cite{digitalopt_micheli1994synthesis}. Its problem formulation remains the same as presented in Eqs.~(\ref{eq:basic_formulation_opt})-(\ref{eq:constrained_formulation_opt}) for analog circuits, with the only difference being that the performance metric $f(\mathbf{x})$ may now represent a digital circuit performance, such as the bit error rate (BER) of a digital system instead of the gain of an operational amplifier. However, unlike its analog counterpart where BO has been extensively used, applying BO to digital circuit optimization appears to be relatively rare. We believe there is an underlying reason for this. Concretely, a digital system is made up of many standard digital cells (e.g., flip-flops, gates, and registers). These cells interact with each other and the overall digital system can be modeled using Boolean logic. Consequently, the performance of the digital system can be easily optimized with respect to all digital cells, frequently with techniques such as linear programming~\cite{digitalopt_burks1993minmax}, integer programming~\cite{digitalopt_liu2002retiming}, and geometric programming~\cite{digitalopt_boyd2005geometric}, while BO is applied occasionally, such as in~\cite{digitalopt_bosystem_li2020exploring,digitalopt_boadder_ma2019cad}. It is important to note that Bayesian optimization is still utilized when the problem focuses on the internal of a digital cell. For example, the authors in~\cite{digitalopt_bosram_wang2017efficient,digitalopt_bosram_zhang2020bayesian,digitalopt_bosram_wang2020efficient} use BO to enhance the yield of an SRAM cell. Nevertheless, it can be debated that their work is essentially equivalent to analog sizing, as the design variables are transistor widths or lengths.

\subsection{Other Problems and Discussions}

As far as we know, addressing schematic-level circuit optimization as described in the previous subsection accounts for the most frequent use of BO in the EDA domain. There are a few works using BO for other circuit problems, which we will summarize in this subsection. Interested readers can directly refer to the cited papers for more details.

The first line of work uses BO to do efficient high-level synthesis (HLS)~\cite{hls_multifid_lo2018multi,hls_accelerator_mehrabi2020bayesian}, logic synthesis~\cite{logicsynthesis_grosnit2022boils,logicsynthesis_zhang2022fast,pd_explain_ustun2019lambda}, and physical (design) synthesis~\cite{pd_explain_geng2023ptpt}. These three concepts are consecutive steps in digital circuit design. High-level synthesis takes an abstract behavioral specification of a digital system and finds a register-transfer level (RTL) structure that realizes the given behavior. Next, logic synthesis takes an RTL description and converts it into a gate-level description. Finally, physical synthesis takes the gate-level description and generates a layout file (e.g., a GDSII file) that an IC foundry can use in manufacturing. Modern EDA tools for HLS, logic synthesis, or physical design synthesis have many parameters, and these parameters (e.g., \emph{directives} in HLS~\cite{hls_multifid_lo2018multi}, Table~I in~\cite{pd_explain_geng2023ptpt}) significantly impact the ultimate performance of the synthesized result. As such, automating the process of tuning these parameters is of great interest, and optimization techniques are being employed for this purpose, among which BO is attractive~\cite{hls_multifid_lo2018multi,hls_accelerator_mehrabi2020bayesian,logicsynthesis_grosnit2022boils,pd_explain_ustun2019lambda,logicsynthesis_zhang2022fast,pd_explain_geng2023ptpt}. See~\cite{hls_reviewbo_ma2019cad} for a review on this topic.

The second line of work uses BO to speed up the design of deep neural network (DNN) accelerators~\cite{accelerator_dnnbo_reagen2017case,accelerator_dnnbo_shi2020learned,accelerator_dnnbo_shi2020using,hls_accelerator_mehrabi2020bayesian,accelerator_tuli2022codebench}. DNNs have gained significant attention due to their successes in various areas at the expense of high computation and memory cost. To improve energy efficiency at deployment, DNN accelerators are promising. The design of a DNN accelerator typically involves hardware (e.g., custom circuits for matrix multiplication) and software (e.g., efficient data-parallel processing algorithm) co-design. The co-design problem is framed as an optimization in the joint space of hardware architectures and software components, and BO has been demonstrated to achieve satisfactory results in this context~\cite{accelerator_dnnbo_reagen2017case,accelerator_dnnbo_shi2020learned,accelerator_dnnbo_shi2020using,hls_accelerator_mehrabi2020bayesian,accelerator_tuli2022codebench}.

BO has also been explored in a few other studies, including efficient automatic test pattern generation~\cite{atpgbo_min2003atspeed,atpgbo_gravagnoli2006automatic}, fast pseudo-transient analysis~\cite{othersbo_pta_xing2022boa}, eye diagram analysis~\cite{othersbo_eye_dolatsara2020worst,others_eye_kiguradze2019bayesian,othersbo_eye_dolatsara2020determining,othersbo_eye_torun2018bayesian}, and layout generation~\cite{bolayout_cicc_chen2021magical,bolayout_liu2020closing}. Because many circuit problems in EDA can be formulated as optimization problems, and the objective functions often do not have analytical forms (i.e., black-box function) and are expensive to evaluate, BO is particularly suitable in these scenarios. However, readers must be alerted that BO might not perform well in a high-dimensional problem. In the current literature, some suggest applying it over continuous domains of fewer than 20
dimensions~\cite{bo_frazier2018tutorial}. Our summary in Table~\ref{tab:bayes_opt_paper} also supports this claim.

In reviewing these papers, we posit the pressing need for a comprehensive and publicly available set of benchmarks in order to facilitate fair comparisons between algorithms. At present, a variety of implementations with minor differences exist as exemplified in Table~\ref{tab:bayes_opt_paper}, and it is unclear which method is the most effective because different research papers utilize disparate examples and evaluation metrics. Furthermore, as various algorithms have distinct parameters that users need to pre-define, it becomes challenging to draw meaningful conclusions. To overcome these issues, we hope that the EDA community collaborates in establishing a set of acknowledged benchmarks. This would enable researchers to compare their algorithms against a standardized set of examples and to draw more credible conclusions. Given the existence of open-source process design kits and EDA tools, we believe that this is a feasible and necessary step toward promoting transparency, collaboration, and advancements in the field.

\section{Future}\label{sec:future}

Finally, we consider a number of potential future research directions in the development and application of Bayesian methods to EDA.

\subsection{Likelihood-free Inference}
In the context of circuit modeling, Bayesian inference often assumes that the form of the likelihood function can be evaluated analytically (e.g., Gaussian or Bernoulli distributions). However, it is important to note that this is simply a model assumption and may not accurately reflect the true nature of the system. Conversely, likelihood-free inference only mandates the capability to sample from the likelihood function, without requiring an analytical form. This type of inference is also referred to as simulator-based inference, as the sampling is performed by executing a simulation. This is exactly the case in a circuit problem where a simulator is available. Hence, we anticipate the potential for likelihood-free inference to be increasingly utilized in circuit-related applications. Current popular likelihood-free inference approaches include approximate Bayesian inference, sequential neural likelihood, and sequential neural posterior, among others. Interested readers can refer to~\cite{lfi_gutmann2016bayesian,lfi_zhang2021unifying,lfi_hermans2020likelihood,lfi_papamakarios2019sequential} for more details.
    
\subsection{Novel Inference on Graphical Models} 
As shown in Table~\ref{tab:bayes_modeling_others}, the use of graphical models (also known as belief networks) is a common approach for representing the relationships between components in a circuit. These graphical models often adopt a tree structure, due to the fact that exact inference is possible through methods such as belief propagation~\cite{bayestheory_bishop2006pattern}. However, this tree structure may not be sufficient for representing the increasingly complicated and entangled relationships between circuit components. Therefore, we advocate for the use of graphical models with loops in circuit problems and for the development of novel methods for performing inference on these models. 
    
\subsection{Hardware/Accelerators for Bayesian Methods} 
We mainly consider Bayesian methods for EDA in this review. However, the reverse direction is also potentially interesting: circuits for implementations of efficient Bayesian methods are also worth exploring. As Bayesian methods (especially Bayesian neural networks, sampling techniques, and inference on graphical models) also account for an important part of modern deep learning, we see the need for specialized hardware for their efficient implementations. In the course of our review, we have seen emerging efforts along this line. The authors in~\cite{accforbayes_osama2016hardware,accforbayes_zhao2016pie} have proposed accelerators that specifically address the challenge of inference on graphical models. Additionally, magnetic tunnel junctions (MTJs) present an opportunity for generating large amounts of true random numbers required for MCMC sampling at a low cost. Integrating MTJs into integrated circuits would provide a promising solution for efficient MCMC sampling. We believe that the development of specialized hardware for Bayesian methods has the potential to significantly improve the efficiency and effectiveness of these methods in a wide range of applications. As such, it represents an exciting area for exploration and investigation.

\subsection{Bayesian Methods for Novel Technologies} 
As we approach the post-Moore era, many novel new technologies are being actively explored. We observe similarities in these emerging areas to needs and approaches used in electronic ICs. An example is silicon photonics, a novel technology utilizing CMOS manufacturing processes to fabricate photonics integrated circuits (PICs). We have already seen a few works exploring Bayesian methods for problems in silicon photonics, ranging from inverse measurement~\cite{bayesfornovel_zhang2022inference}, optimization~\cite{bo_siliconphotonics_gao2022automatic}, machine learning~\cite{bayesfornovel_sarantoglou2022bayesian}, and quantum applications~\cite{bayesfornovel_paesani2017experimental}.

\medskip

In conclusion, the application of Bayesian methods in EDA is and will remain a growing field with great potential, due to the continuing development of electronic systems and progress in statistics. 
\section{Conclusion}\label{sec:conclusion}

This comprehensive review of Bayesian methods in EDA presents the basic concepts and published literature within the field. We organize the literature into two main categories: (i)~modeling, diagnosis, and analysis; and (ii)~optimization, and provide a detailed review of each category. Finally, we identify several promising avenues for future research and open questions that warrant further investigation. By highlighting the benefits and potential of Bayesian methods, we hope to inspire their wider adoption and application in solving various challenges in EDA and future novel technologies.

\bibliography{source}
\bibliographystyle{IEEEtran}
\end{document}